\documentclass[11pt]{article}

\usepackage{graphicx}
\usepackage{latexsym}
\usepackage{enumerate}
\usepackage{amsmath}
\usepackage{amsfonts}
\usepackage{amssymb}
\usepackage{amsthm}
\usepackage{amsopn}
\usepackage{amstext}
\usepackage{amscd}

\newtheorem{theorem}{Theorem}[section]
\newtheorem{proposition}[theorem]{Proposition}
\newtheorem{corollary}[theorem]{Corollary}
\newtheorem{example}[theorem]{Example}

\begin{document}

\newcommand{\PFN}{I_F^N}
\newcommand{\PGN}{I_G^N}
\newcommand{\PHN}{I_H^N}
\newcommand{\PGHN}{I_{G+H}^N}
\newcommand{\bigabs}[1]{\big\lvert#1\big\rvert}
\newcommand{\Bigabs}[1]{\Big\lvert#1\Big\rvert}
\newcommand{\bbZ}{\mathbb{Z}}
\newcommand{\bbR}{\mathbb{R}}
\newcommand{\bbN}{\mathbb{N}}
\newcommand{\zkone}{z_k^{(1)}}
\newcommand{\zktwo}{z_k^{(2)}}
\newcommand{\ckone}{c_k^{(1)}}
\newcommand{\cktwo}{c_k^{(2)}}
\newcommand{\skone}{s_k^{(1)}}
\newcommand{\sktwo}{s_k^{(2)}}
\newcommand{\Cnl}{\mathcal{C}_0}
\newcommand{\D}{\mathcal{D}}
\newcommand{\tldFone}{G}
\newcommand{\tldfone}{g}
\newcommand{\transp}{{^\text{T}}}
\newcommand{\Fourk}{\mathcal{F}_k}
\newcommand{\ckG}{c_{k,G}}
\newcommand{\ckH}{c_{k,H}}
\newcommand{\skG}{s_{k,G}}
\newcommand{\skH}{s_{k,H}}
\newcommand{\abs}[1]{\left|#1\right|}

%

\title{A METHOD OF TREND EXTRACTION USING\linebreak SINGULAR SPECTRUM ANALYSIS}
\author{Theodore Alexandrov \\
                      Center for Industrial Mathematics,
                       University of Bremen,\\
                       Germany
                       \ (theodore@math.uni-bremen.de)}


\maketitle

\begin{abstract}
The paper presents a new method of trend extraction in the framework
of the Singular Spectrum Analysis (SSA) approach. This method is easy to use,
does not need specification of models of time series and trend,
allows to extract trend in the presence
of noise and oscillations and has only two parameters (besides basic SSA parameter called window length).
One parameter manages scale of the extracted trend and another is a method specific threshold value.
We propose procedures for the choice of the parameters.
The presented method is evaluated on a simulated time series with a polynomial trend 
and an oscillating component with unknown period
and on the seasonally adjusted monthly data of unemployment level in Alaska
for the period 1976/01-2006/09.


\end{abstract}


\section{INTRODUCTION}
Trend extraction is an important task in applied time series analysis,
in particular in economics and engineering. We present a new
method of trend extraction in the framework of the Singular Spectrum Analysis approach.

Trend is usually defined as {a smooth additive component containing information
about time series global change}. This definition is rather vague (which type of smoothness is used?
which kind of information is contained in the trend?). It may sound strange, but there is no more precise
definition of the trend accepted by the majority of researchers and practitioners.
Each approach to trend extraction defines trend with respect to the mathematical tools used 
(e.g. using Fourier transformation or derivatives).
Thus in the corresponding literature one can find various specific definitions of the trend.
For further discussion on trend issues we refer to~\cite{TRreview08}.

Singular Spectrum Analysis (SSA) is a general approach to time series analysis and forecast.
Algorithm of SSA is similar to that of Principal Components Analysis (PCA) of multivariate data.
In contrast to PCA which is applied to a matrix, SSA is applied to a time series and provides 
a representation of the given time series in terms of eigenvalues and eigenvectors of a matrix 
made of the time series.
The basic idea of SSA has been proposed by~\cite{BK86} for dimension calculation and reconstruction 
of attractors of dynamical systems, see historical reviews in~\cite{ASMCTS02} and in~\cite{GNZ01}.
In this paper we mostly follow the notations of~\cite{GNZ01}.

SSA can be used for a wide range of tasks: trend or quasi-periodic component detection and extraction,
denoising, forecasting, change-point detection. 
The present bibliography on SSA includes two monographes,
several book chapters, and over a hundred papers. For more details see references at
the website SSAwiki: {http://www.math.uni-bremen.de/$\sim$theodore/ssawiki}.

The method presented in this paper has been first proposed in~\cite{AlexGol_workshop05}
and is studied in detail in the author's unpublished Ph.D. thesis \cite{PhDthesis}
available only in Russian at {http://www.pdmi.ras.ru/$\sim$theo/autossa}.

The proposed method is easy to use (has only two parameters), does not need specification of models of time series and trend,
allows one to specify desired trend scale, and extracts trend in the presence of noise and oscillations.

The outline of this paper is as follows. Section~\ref{sec:SSA} introduces SSA,
formulates properties of trends in SSA and presents the already existing 
methods of trend extraction in SSA. Section~\ref{sec:method} proposes 
our method of trend extraction. In section~\ref{sec:omega0} we discuss
the frequency properties of additive components of a time series and 
present our procedure for the choice of first parameter of the method,
a low-frequency boundary. Section~\ref{sec:param_C0} starts with investigation 
of the role of the second method parameter, the low-frequency contribution,
based on a simulation example. Then we propose a heuristic strategy 
for the choice of this parameter.
In section~\ref{sec:example}, applications of the proposed method to a simulated
time series with a polynomial trend and oscillations and on the unemployment level in Alaska 
are considered.
Finally, section~\ref{sec:conclusions} offers conclusions.

\section{SINGULAR SPECTRUM ANALYSIS}
\label{sec:SSA}
Let us have a time series $F=(f_0,\ldots,f_{N-1})$, $f_n\in\mathbb{R}$, of length $N$,
and we are looking for some specific additive component of $F$ (e.g. a trend).
The central idea of SSA is to embed $F$ into
high-dimensional euclidean space, then find a subspace corresponding to the sought-for
component and, finally, reconstruct a time series component corresponding to this subspace.
The choice of the subspace is a crucial question in SSA.
The basic SSA algorithm consists of {decomposition} of a time series and 
{reconstruction} of a desired additive component. These two steps are summarized below;
for a detailed description, see page 16 of~\cite{GNZ01}.

\textbf{Decomposition.} The decomposition takes a time series of length $N$ and comes up 
with an $L \times K$ matrix. This stage starts by defining a parameter $L$ ($1<L<N$), 
called the {window length}, and constructing the so-called trajectory matrix 
${\bf X}\in\mathbb{R}^{L\times K}$, $K=N-L+1$,
with stepwise taken portions of the original time series $F$ as columns:
\begin{equation}
\label{eq:trajmatrixconstruction}
F=(f_0,\ldots,f_{N-1}) \rightarrow {\bf X}=[X_1:\ldots:X_K], ~~ X_j=(f_{j-1},\ldots,f_{j+L-2}){^\text{T}}.
\end{equation}
Note that ${\bf X}$ is a Hankel matrix and (\ref{eq:trajmatrixconstruction}) defines one-to-one
correspondence between series of length $N$ and Hankel matrices of size $L\times K$.
Then Singular Value Decomposition (SVD) of ${\bf X}$ is applied, where $j$-th component of SVD is
specified by $j$-th eigenvalue $\lambda_j$ and eigenvector $U_j$ of ${\bf X}{\bf X}{^\text{T}}$:
\begin{equation*}
{\bf X} = \sum_{j=1}^d \sqrt{\lambda_j} U_j V_j{^\text{T}}, \quad 
V_j={\bf X}{^\text{T}} U_j\Bigl/\sqrt{\lambda_j}, \quad d=\max \{j: ~ \lambda_j>0\}.
\end{equation*}
Since the matrix ${\bf X}{\bf X}{^\text{T}}$ is positive-definite,
their eigenvalues $\lambda_j$ are positive.
The SVD components are numbered in the decreasing order of eigenvalues $\lambda_j$.
We define $j$-th Empirical Orthogonal Function (EOF) as the sequence of elements
of the $j$-th eigenvector $U_j$. The triple $(\sqrt{\lambda_j},U_j,V_j)$ is called
$j$-th eigentriple, $\sqrt{\lambda_j}$ is called the $j$-th singular value, $U_j$ is the $j$-th
left singular vector and $V_j$ is the $j$-th right singular vector.

\textbf{Reconstruction.} Reconstruction goes from an $L \times K$ matrix into
a time series of length $N$. This stage combines (i) selection of a subgroup $\mathcal{J}\subset\{1,\ldots,L\}$ of SVD components;
(ii) hankelization (averaging along entries with indices $i+j=const$) of the $L \times K$ matrix 
from the selected $\mathcal{J}$ components of the SVD; (iii) reconstruction of a time series 
component of length $N$ from the Hankel matrix by the mentioned one-to-one correspondence
(like in (\ref{eq:trajmatrixconstruction}) but in the reverse direction, see below the 
exact formulae). The result of the reconstruction stage is a time series additive component:
\begin{equation*}
        {\bf X}_{\mathcal{J}} = \sum_{j\in\mathcal{J}} \sqrt{\lambda_j} U_j V_j{^\text{T}} ~ \rightarrow ~
        G=(g_0, \ldots, g_{N-1}).
\end{equation*}

For the sake of brevity, let us describe the hankelization of the matrix ${\bf X}_{\mathcal{J}}$ 
and the subsequent reconstruction of a time series component $G$ as being applied to 
a matrix ${\bf Y}=\bigl\{ y_{ij} \bigr\}_{i,j=1}^{i=L,j=K}$ as it is introduced in~\cite{GNZ01}.
First we introduce ${L}^* = \min\{L,K\}$, ${K}^* = \max\{L,K\}$ and define an $L^* \times K^*$ 
matrix ${\bf Y}^*$ as given by ${\bf Y}^*={\bf Y}$ if $L\leqslant K$ and ${\bf Y}^*={\bf Y}\transp$ if $L>K$.
Then the elements of the time series $G=(g_0,\ldots,g_{N-1})$ formed from the matrix $\bf Y$ are calculated 
by averaging along cross-diagonals of matrix ${\bf Y}^*$ as
\begin{equation}
\label{eq:ssa.diagaver}
         g_n=\begin{cases}
        \frac{1}{n+1}\sum\limits_{m=1}^{n+1}{y}^*_{m,n-m+2},\quad
                0 \leqslant n < {L}^*-1, \\

        \frac{1}{{L}^*}\sum\limits_{m=1}^{{L}^*}{y}^*_{m,n-m+2},\quad
                {L}^*-1 \leqslant n < {K}^*, \\

        \frac{1}{N-n}\sum\limits_{m=n-{K}^*+2}^{N-{K}^*+1}{y}^*_{m,n-m+2},\quad
                {K}^* \leqslant n < N.
        \end{cases}
\end{equation}

Changing the window length parameter and, what is more important, the subgroup $\mathcal{J}$ 
of SVD components used for reconstruction, one can change the output time series $G$. 
In the problem of trend extraction, we are looking for $G$ approximating a trend of a time series.
Thus, the trend extraction problem in SSA is reduced to (i) the choice of a window length $L$ 
used for decomposition and
(ii) the selection of a subgroup $\mathcal{J}$ of SVD components used for reconstruction.
The first problem is thoroughly discussed in section 1.6~of~\cite{GNZ01}.
In this paper, we propose a solution for the second problem.

Note that for the reconstruction of a time series component,
SSA considers the whole time series, as its algorithm uses SVD of the trajectory matrix
built from all parts of the time series.  Therefore, SSA is not a local method in contrast to
a linear filtering or wavelet methods. On the other hand, this property makes SSA robust 
to outliers, see~\cite{GNZ01} for more details.

An essential disadvantage of SSA is its computational complexity for the calculation of SVD.
This shortcoming can be reduced by using modern~\cite{DrmVese05} and parallel algorithms for SVD.
Moreover, for trend revision in case of receiving new data points, a computationally attractive
algorithm of~\cite{GuEis93} for updating SVD can be used.

It is worth to mention here that the similar ideas of using SVD of the trajectory matrix 
have been proposed in other areas, e.g. in signal extraction in oceanology~\cite{Co78} 
and estimation of parameters of damped complex exponential signals~\cite{KT80}.

\subsection{Trend in SSA}
\label{sec:trendinSSA}
SSA  is a nonparametric approach which does not need {a priori} specification of models of time series and trend,
neither deterministic nor stochastic ones.
The classes of trends and residuals which can be successfully separated
by SSA are characterized as follows.

First, since we extract any trend by selecting a subgroup of all $d$ SVD components, this trend
should generate less than $d$ SVD components. For an infinite time series, a class of
such trends coincides with the class of time series governed by
finite difference equations~\cite{GNZ01}. This class
can be described explicitly as linear combinations of products of polynomials, exponentials and 
sines~\cite{Buch94}. An element of this class suits well for representation of a smooth and slow varying trend.

Second, a residual should belong to a class of time series which can be separated from a trend.
The separability theory due to~\cite{Nekrutkin96} helps us determine this class. 
In \cite{Nekrutkin96} it was proved
that (i) any deterministic function can be asymptotically separated from any ergodic
stochastic noise as the time series length and window length tend to infinity;
(ii) under some conditions any trend can be separated from any quasi-periodic component,
see also~\cite{GNZ01}.
These properties of SSA make this approach feasible for trend extraction in the presence of noise and
quasi-periodic oscillating components.

Finally, as trend is a smooth and slow varying time series component, it generates SVD components
with smooth and slow varying EOFs. Eigenvectors represent an orthonormal basis of a trajectory
vector space spanned on the columns of trajectory matrix. Thus each EOF
is a linear combination of portions of the corresponding time series and inherits
its global smoothness properties. This idea is considered in detail in~\cite{GNZ01}
for the cases of polynomial and exponential trends.

\subsection{Existing methods of trend extraction in SSA}
A naive approach to trend extraction in SSA is to reconstruct a trend from several first SVD components.
Despite its simplicity, this approach works in many real-life cases
for the following reason.
An eigenvalue represents a contribution of the corresponding SVD component
into the form of the time series, see section 1.6 of~\cite{GNZ01}. 
Since a trend usually characterizes the shape of a time series, its eigenvalues are larger than the other ones,
that implies small order numbers of the trend SVD components.
However, the selection procedure fails when the values of a trend are small enough as compared
with a residual, or when a trend has a complicated structure (e.g. a high-order polynomial)
and is characterized by many (not only by the first ones) SVD components.

A smarter way of selecting trend SVD components is to choose the components with
smooth and slow varying EOFs (we have explained this fact above). At present, there exist only
one parametric method of~\cite{VYG92} which follows this approach.
In \cite{VYG92} it was proposed using the Kendall correlation coefficient for testing
for monotonic growth of an EOF. Unfortunately, this method is far from perfect since 
it is not possible to establish which kinds of trend can be extracted by its means. 
This method seems to be aimed at extraction of monotonic trends because their EOFs are usually monotonic. 
However, even a monotonic trend can produce non-monotonic EOF, especially in case of noisy
observations. An example could be a linear trend which generates a linear and a constant EOFs. 
If there is a noise or another time series component added, then this component is often mixed with 
trend components corrupting its EOFs. Then, even in case of very small corruption, 
the constant EOF can be highly non monotonic. Naturally, the method using the Kendall correlation 
coefficient does not suit for non monotonic trends producing non monotonic EOFs. 
For example, a polynomial of low order which is often used for trend modelling usually produces 
non monotonic EOFs, for details see e.g.~\cite{GNZ01}.

\section{PROPOSED METHOD FOR TREND EXTRACTION}
\label{sec:method}
In this section, we present our method of trend extraction. First, following~\cite{GNZ01}, 
we introduce the periodogram of a time series. 

Let us consider the Fourier representation of the elements of a time series $X$ of length $N$, 
$X=(x_0,\ldots,x_{N-1})$, see e.g. section 7.3 of~\cite{Chatf96}:
\begin{equation*}
\label{def:Fourier_expansion}
x_n=c_0+\sum_{1 \leqslant k \leqslant \frac{N-1}{2}}\bigl(
        c_k\cos(2\pi n k/N) + s_k\sin(2\pi n k/N)
\bigr)+(-1)^n c_{N/2},
\end{equation*}
where $k\in\mathbb{N}$, $0\leqslant n \leqslant N-1$, and $c_{N/2}=0$ if $N$ is an odd number. 
Then the periodogram of $X$ at the frequencies $\omega\in\{k/N\}_{k=0}^{\lfloor N/2 \rfloor}$ 
is defined as
\begin{equation}
\label{eq:periodogram.Fourier}
I_X^N(k/N) = \frac{N}{2}
        \begin{cases}
                2c_0^2, \quad k=0, \\
                c_k^2 + s_k^2, \quad 0< k < N/2, \\
                2c_{N/2}^2, \quad \text{if $N$ an even number and $k=N/2$}.
        \end{cases}
\end{equation}

Note that this periodogram is different from the periodogram usually used in 
spectral analysis, see e.g.~\cite{Brill2001} or \cite{Chatf96}. 
To show this difference, let us denote the $k$-th element of the discrete Fourier transform of $X$ as 
\begin{equation*}
\Fourk(X) = \sum_{n=0}^{N-1} e^{-i2\pi n k/N} x_n,
\end{equation*}
then the periodogram $I_{X}^N(\omega)$ at the frequencies $\omega\in\{k/N\}_{k=0}^{\lfloor N/2 \rfloor}$ 
is calculated as
\begin{equation*}
I_{X}^N(k/N) = \frac{1}{N}
\begin{cases}
   2 \left| \Fourk(X) \right|^2, \quad \text{if}~0 < k <N/2, \\
    \left| \Fourk(X) \right|^2, \quad \text{if}~k=0~\text{or $N$ is even and}~k=N/2.
\end{cases}
\end{equation*}
One can see that in addition to the normalization different from that in~\cite{Brill2001} and~\cite{Chatf96},
the values for frequencies in the interval $(0,0.5)$ are multiplied by two. 
This is done to ensure the following property:
\begin{equation}
\label{eq:L2norm_periodogram}
\left|\left| X \right|\right|_2^2 = \sum_{n=0}^{N-1} x_n^2 = \sum_{k=0}^{\lfloor N/2\rfloor} I_{X}^N(k/N).
\end{equation}

Let us introduce the cumulative contribution of the frequencies $[0,\omega]$ as
$\pi_{X}^N(\omega)= \sum_{k: 0\leq k/N \leq\omega}I_X^N(k/N)$, $\omega\in [0,0.5]$.
Then, for a given $\omega_0\in(0,0.5)$, we define the contribution of low frequencies from the interval $[0,\omega_0]$ to  $X\in\mathbb{R}^N$ as
\begin{equation}
\label{eq:C_Xomega}
\mathcal{C}(X, \omega_0)=\pi_{X}^N(\omega_0)/\pi_{X}^N(0.5).
\end{equation}
Then, given parameters $\omega_0\in(0,0.5)$ and $\mathcal{C}_0\in[0,1]$, we propose to select
those SVD components whose eigenvectors satisfy the following criterion:
\begin{equation}
\label{eq:method.selectioncrit}
\mathcal{C}(U_j, \omega_0) \geqslant \mathcal{C}_0,
\end{equation}
where $U_j$ is the corresponding $j$-th eigenvector. One may interpret this
method as selection of SVD components with EOFs mostly  characterized
by low-frequency fluctuations. It is worth noting here that when we apply
$\mathcal{C}$, $\pi$ or $I$ (defined above for a time series) to a vector, they are simply applied
to a series of elements of the vector.

Having the trend SVD components selected using (\ref{eq:method.selectioncrit}), 
one reconstructs the trend according to section~\ref{sec:SSA}. The question
is how to select $\omega_0$ and how to define the threshold $\mathcal{C}_0$.
These issues are discussed in sections \ref{sec:omega0} and \ref{sec:param_C0},
respectively.

\section{THE LOW-FREQUENCY BOUNDARY $\omega_0$}
\label{sec:omega0}
The low-frequency boundary $\omega_0$ manages the scale of the extracted trend:
the lower is $\omega_0$, the slower varies the extracted trend. Selection of 
$\omega_0$ can be done a priori based on additional information about the data 
thus prespecifying the desired scale of the trend. 

For example, if we assume to have a quasi-periodic component with known period $T$, 
then we should select $\omega_0 < 1/T$ in order not to include this component
in the trend. For extraction of a trend of monthly data
with possible seasonal oscillations of period 12, we suggest to select $\omega_0< 1/12$,
e.g.~$\omega_0=0.075$.

In this paper we also propose a method of selection of $\omega_0$ considering 
a time series periodogram.
Since a trend is a slow varying component, its periodogram has large values
close to zero frequency and small values for other frequencies. The problem of selecting
$\omega_0$ is the problem of finding such a low-frequency value that 
the frequencies corresponding to the large trend periodogram values are inside
the interval $[0,\omega_0]$. At the same time, $\omega_0$ cannot be too large
because then an oscillating component with a frequency less than $\omega_0$ can 
be included in the trend produced. Considering the periodogram of a trend,
we could find the proper value of $\omega_0$ but for a given time series its
trend is unknown.

What we propose is to choose $\omega_0$ based on the periodogram of the original 
time series. The following proposition substantiates this approach.

\begin{proposition}
\label{prop:LF.Pi(G+H)}
Let us have two time series $G=(g_0,\ldots,g_{N-1})$ and $H=(h_0,\ldots,h_{N-1})$ of length $N$,
then for each $k$: $0\leq k \leq \lfloor N/2 \rfloor$ the following inequality holds:
\begin{equation}
\label{eq:LF.PI(G+H)-PI(G)-PI(H)}
\big\lvert I_{G+H}^N(k/N) - I_G^N(k/N) - I_H^N(k/N) \big\rvert \leqslant
2\sqrt{I_G^N(k/N)I_H^N(k/N)}.
\end{equation}
\end{proposition}
\begin{proof}[Proof of Proposition~\ref{prop:LF.Pi(G+H)}]
Let us first consider the case when $0<k<N/2$. We denote as $c_{k,X}$ and $s_{k,X}$ 
the coefficients of Fourier representation of a time series $X$ used in the periodogram
definition (\ref{eq:periodogram.Fourier}). Then, by this definition,
\begin{equation*}
\begin{array}{c}
\displaystyle \PGHN(k/N)-\PGN(k/N)-\PHN(k/N) =~~~~~~~~~~~~~~~~~~~~~ \\
\vspace{-3mm}\\
~~~~~~~~\displaystyle =\frac{N}{2} \left( c_{k,G+H}^2 + s_{k,G+H}^2 - \ckG^2 - \skG^2 - \ckH^2 - \skH^2 \right).
\end{array}
\end{equation*}
Since $c_{k,G+H}=\frac{2}{N} \Re\Fourk(G+H)=\ckG+\ckH$ (where $\Re z$ denotes an imaginary part of a complex number $z$) 
and, analogously, $s_{k,G+H}=\skG+\skH$, we have 
\begin{equation}
\label{eq:IGH-IG-IH}
\PGHN(k/N)-\PGN(k/N)-\PHN(k/N) = N \left( \ckG\ckH + \skH\skH \right).
\end{equation}
Let us consider the periodograms multiplication used in the right part of (\ref{eq:LF.PI(G+H)-PI(G)-PI(H)}):
\begin{equation}
\label{eq:IG*IG_1}
\PGN(k/N)\PHN(k/N) = \frac{N^2}{4} \left(\ckG^2+\skG^2 \right)\left(\ckH^2+\skH^2 \right).
\end{equation}
Since for all real $a,b,c,$ and $d$ it holds that $(a^2+b^2)(c^2+d^2) = (\abs{ac}+\abs{bd})^2+(\abs{ad}-\abs{bc})^2$,
then
\begin{equation}
\label{eq:IG*IH}
\PGN(k/N)\PHN(k/N) = \frac{N^2}{4} \left(\abs{\ckG\ckH}+\abs{\skG\skH} \right)^2 + \left(\abs{\ckG\skH}-\abs{\ckH\skG} \right)^2.
\end{equation}
Finally, taking the square of (\ref{eq:IGH-IG-IH}), dividing it by four and taking into account~(\ref{eq:IG*IH}), we have
\begin{eqnarray*}
   \displaystyle \frac{1}{4}\bigl( \PGHN(k/N)-\PGN(k/N)-\PHN(k/N) \bigr)^2 = ~~~~~~~~~~~~~~~~~~~~~~~~~~~~~~~~~~~~~~~~~  \\
   \displaystyle = \frac{N^2}{4} \left(\ckG\ckH+\skG\skH \right)^2 \leqslant  \frac{N^2}{4} \left(\abs{\ckG\ckH}+\abs{\skG\skH} \right)^2 \leqslant ~~~~~~~~~~~ \\
   \displaystyle \leqslant \frac{N^2}{4} \left(\abs{\ckG\ckH}+\abs{\skG\skH} \right)^2 + \left(\abs{\ckG\skH}-\abs{\ckH\skG} \right)^2 = ~~~~~ \\
   \displaystyle = \PGN(k/N)\PHN(k/N) \\
\end{eqnarray*}
and the inequality in (\ref{eq:LF.PI(G+H)-PI(G)-PI(H)}) holds $0<k<N/2$.

Second, we consider the case when $k=0$ or $k=N/2$. Again, by the definition of the periodogram
\begin{equation*}
   2\sqrt{\PGN(k/N)\PHN(k/N)} = 2 \sqrt{N^2 \ckG^2 \ckH^2} = 2 N \left| \ckG \ckH \right|.
\end{equation*}
At the same time, 
\begin{equation*}
   \bigl| \PGHN(k/N)-\PGN(k/N)-\PHN(k/N) \bigr| = N \left| c_{k,G+H}^2 - \ckG^2 - \ckH^2  \right| = N \left| 2 \ckG\ckH \right|
\end{equation*}
which leads for $k=0$ or $k=N/2$ to 
\begin{equation*}
   \bigl| \PGHN(k/N)-\PGN(k/N)-\PHN(k/N) \bigr| = 2\sqrt{\PGN(k/N)\PHN(k/N)}
\end{equation*}
and the result in (\ref{eq:LF.PI(G+H)-PI(G)-PI(H)}) holds with equality.

\end{proof}

\begin{corollary}
\label{cor:indepsupports}
Let us define for a time series $F$ of length $N$ the frequency support of the periodogram $I_F^N$ as 
a subset of frequencies $\{k/N\}_{k=0}^{\lfloor N/2 \rfloor}$ such that $I_F^N(k'/N)> 0$ for $k'/N$
from this subset.
If the frequency supports of two time series $G$ and $H$ are disjoint then $\PGHN(k/N) = \PGN(k/N) + \PHN(k/N)$
\end{corollary}

Let us demonstrate that when supports of periodograms of time series $G$ and $H$ are nearly disjoint,
the periodogram of the sum $G+H$ is close to the sum of their periodograms.

The fact that the periodograms of $G$ and $H$ are very different at $k/N$ 
can be expressed as 
\begin{equation*}
\PGN(k/N)/\PHN(k/N) = d \gg 1,
\end{equation*}
since without loss of generality we can assume $\PGN(k/N)>\PHN(k/N)$. 
Then using Proposition~\ref{prop:LF.Pi(G+H)} we have that
\begin{equation*}
\begin{array}{c}
\displaystyle \big\lvert \PGHN(k/N) - \PGN(k/N) - \PHN(k/N) \big\rvert \leqslant ~~~~~~~~~~~~~~~~~~~~~ \\
\vspace{-3mm}\\
~~~~~~~~~~~~ \displaystyle  \leqslant 2\sqrt{\PGN(k/N)\PHN(k/N)} = \frac{2}{\sqrt{d}} \PGN(k/N) \ll \PGN(k/N),
\end{array}
\end{equation*}
that means that the difference $\bigl\lvert \PGHN(k/N) - \PGN(k/N) - \PHN(k/N) \bigr\rvert$ 
is significantly smaller than the value of the largest periodogram (of $\PGN$, $\PHN$) at the point $k/N$.

In many applications, the given time series can be modelled as made of a trend
with large periodogram values at low-frequency interval $[0,\omega_0]$,
oscillations with periods smaller than $1/\omega_0$,
and noise whose frequency contribution spreads over all the frequencies $[0,0.5]$ but is relatively small.
In this case the periodogram supports of the trend and the residual can be considered as nearly disjoint.
Therefore, from Corollary~\ref{cor:indepsupports}, we conclude that 
the periodogram of the time series is approximately equal to the sum of the periodograms
of the trend, oscillations and noise.

For a time series $X$ of length $N$, we propose to select the value of the parameter $\omega_0$ 
according to the following rule:
\begin{equation}
\omega_0 = \max_{k/N, 0\leqslant k \leqslant N/2} \bigl\{k/N:  I_X^N(0),\ldots,I_X^N(k/N) < M_X^N \bigr\}, 
\label{eq:omega0_rule}
\end{equation}
where $M_X^N$ is the median of the values of periodogram of $X$. The modelling of a time series
as a sum of a trend, oscillations and a noise (let us suppose to have a normal noise) 
motivates this rule as follows. Since the frequency 
supports of the trend and oscillating components do not overlap, only the values of the noise 
periodogram can mix with the values of the trend periodogram. First, the values of the noise periodogram 
for neighboring ordinates are asymptotically independent (see e.g. section 7.3.2 of~\cite{Chatf96}). 
Second, supposing a relatively long time series and narrow frequency supports of 
trend and oscillating components, the median of values of the time series periodogram gives an estimation
of the median of the values of the noise periodogram. Since a trend is supposed to have large
contribution to the shape of the time series (i.e. a large $L_2$-norm) compared to the noise
and its frequency support is quite narrow compared to the whole interval $[0,0.5]$,
its periodogram values are relatively larger than the median of the noise periodogram values due to~(\ref{eq:L2norm_periodogram}). 
Therefore, the condition used in (\ref{eq:omega0_rule}) is fulfilled only 
for such a frequency $\omega_0$ that 
the trend periodogram values is close to zero (outside the trend frequency interval). 
Large noise periodogram values in this frequency region can lead to selecting larger than necessary $\omega_0$. 
But remember that we compare the periodogram values with their median and 
the noise periodogram values are independent (asymptotically). Hence, with probability approximately 
equal to $1-0.5^m$ (e.g. this value is equal to 0.9375 for $m=4$) 
we select the $m$-th point (of the grid $\{k/N\}$) located to the right side of the trend frequency
interval (where the trend peridogram values are larger then the noised periodogram median).

Note that the lengths $N$ of the time series and $L$ of eigenvector are different ($L<N$)
which causes different resolution of their periodograms. Having estimated $\omega_0$ after 
consideration of the periodogram of the original time series, one should select 
\begin{equation}
\label{eq:w0_LN}
\omega_0'=\lceil L \omega_0 \rceil / L.
\end{equation}

\paragraph{Dependence of $\omega_0$ on the time series resolution.}
Let us define the resolution $\rho$ of the original time series as $\rho=(\tau_{n+1}-\tau_n)^{-1}$, where
$\tau_n$ is the time of $n$-th measurement. If one have estimated $\omega_0$ for the data with resolution $\rho$
and there comes the same data but measured with higher resolution $\rho'=m\rho$ ($m\in\mathbb{R}$) 
thus increasing the data length in $m$ times, 
then in order to extract the same trend, one should take the new threshold value $\omega_0'=\omega_0/m$.
In a similar manner, after decimation of the data reducing the resolution in $m$ times, 
the value $\omega_0'=m\omega_0$ should be taken.

\begin{example}[The choice of $\omega_0$ for a noised exponential trend]
\label{ex:omega0_exptrend_noise}
Let us consider an example of selection of the treshold $\omega_0$ for an exponential trend and 
a white Gaussian noise which also demonstrates Proposition~\ref{prop:LF.Pi(G+H)}.
 Let the time series $F=G+H$ be of length $N=120$, 
where the components $G$ and $H$ are defined as $g_n=A e^{0.01 n}$, $h_n=B \varepsilon_n$,
$\varepsilon_n \sim iidN(0,1)$ and $A,B$ are selected so that 
$||G||^2=||H||^2=\sum_{n=0}^{N-1}g_n=\sum_{n=0}^{N-1}h_n=1$. The normalization is done to ensure that
$\sum_{k=0}^{60} \PGN(k/N)=\sum_{k=0}^{60} \PHN(k/N)=1$. Figure~\ref{fig:LF.Pi(G+H)}
shows a) the simulated time series $F$, b) its components, c) the periodograms of the components, 
d) the periodograms zoomed together with a line corresponding to the median of the noise periodogram
values equal to 0.0126, e) the periodogram $\PFN$ of $F$ and a kind of ``confidence'' interval of 
its estimation $\PGN+\PHN$ calculated according Proposition~\ref{prop:LF.Pi(G+H)} and 
a line corresponding to the median $M_F^{120}$ of the time series periodogram values 
(used for estimating $\omega_0$), and 
f) the discrepancy, the difference between $\PFN$ and $\PGN+\PHN$ together with the values of this
difference estimated in the right side of~(\ref{eq:LF.PI(G+H)-PI(G)-PI(H)}). Note tha the median of the 
periodogram values of $F$ is equal to 0.0141, which is close to the median of the noise periodogram 
values equal to 0.0126.
The value of $\omega_0$
estimated according to the proposed rule~(\ref{eq:omega0_rule}) is equal to $6/120=0.05$.
\vspace*{5mm}
\begin{figure}[ht]
\begin{center}
\begin{minipage}{1.1\textwidth}
	\begin{minipage}{0.3\textwidth}
	\centerline{\textsf{\small {\bf a)} Original time series}}
	\centerline{\textsf{\small $F=G+H$}}
	\centerline{\includegraphics[width=0.9\textwidth]{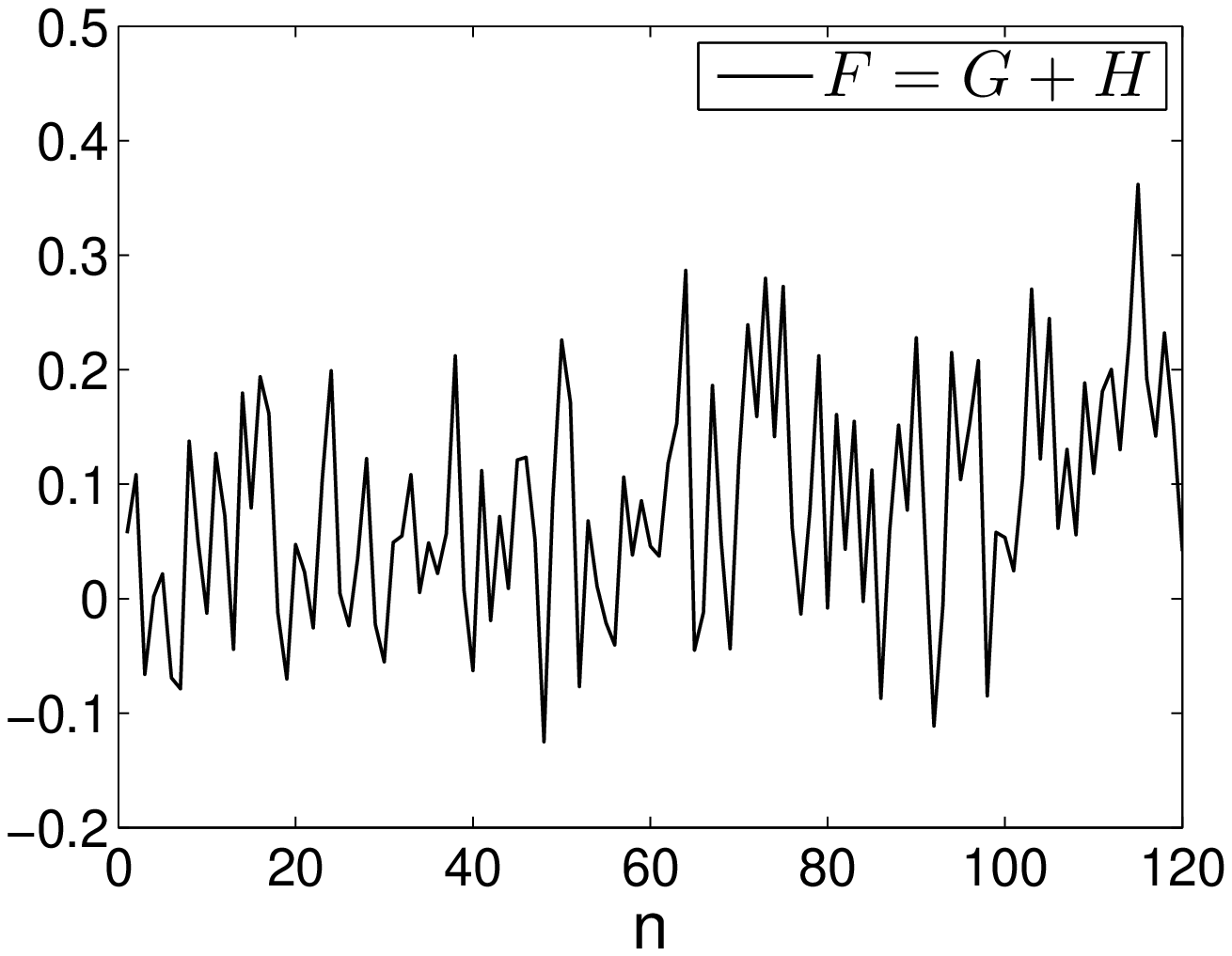}}
	\end{minipage}
	\begin{minipage}{0.3\textwidth}
	\centerline{\textsf{\small {\bf b)} Time series components}}
	\centerline{\textsf{\small $G$ and $H$}}
	\centerline{\includegraphics[width=0.9\textwidth]{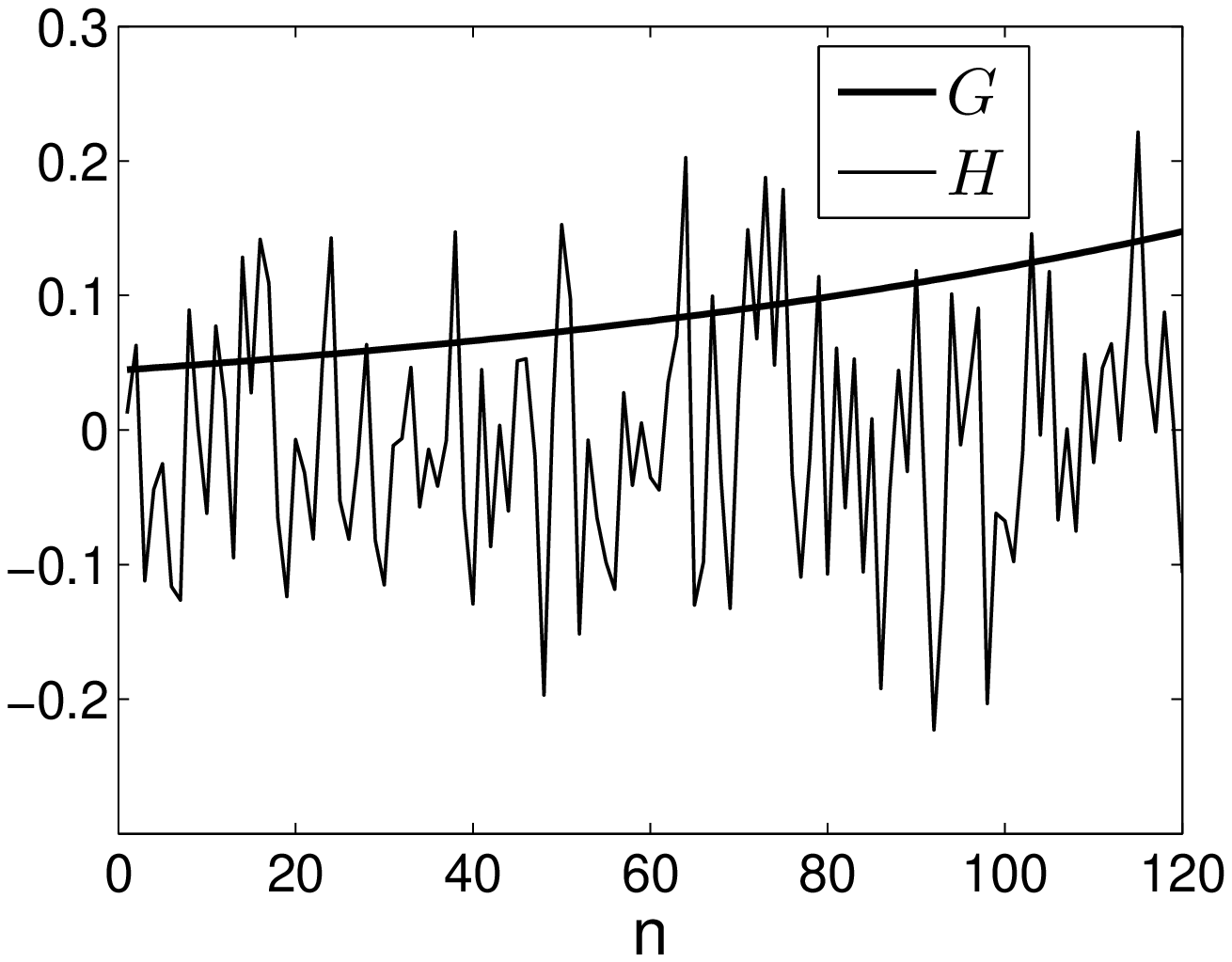}}
	\end{minipage}
	\begin{minipage}{0.3\textwidth}
	\centerline{\textsf{\small {\bf c)} Periodograms of}}
	\centerline{\textsf{\small components $G$ and $H$}}
	\centerline{\includegraphics[width=0.9\textwidth]{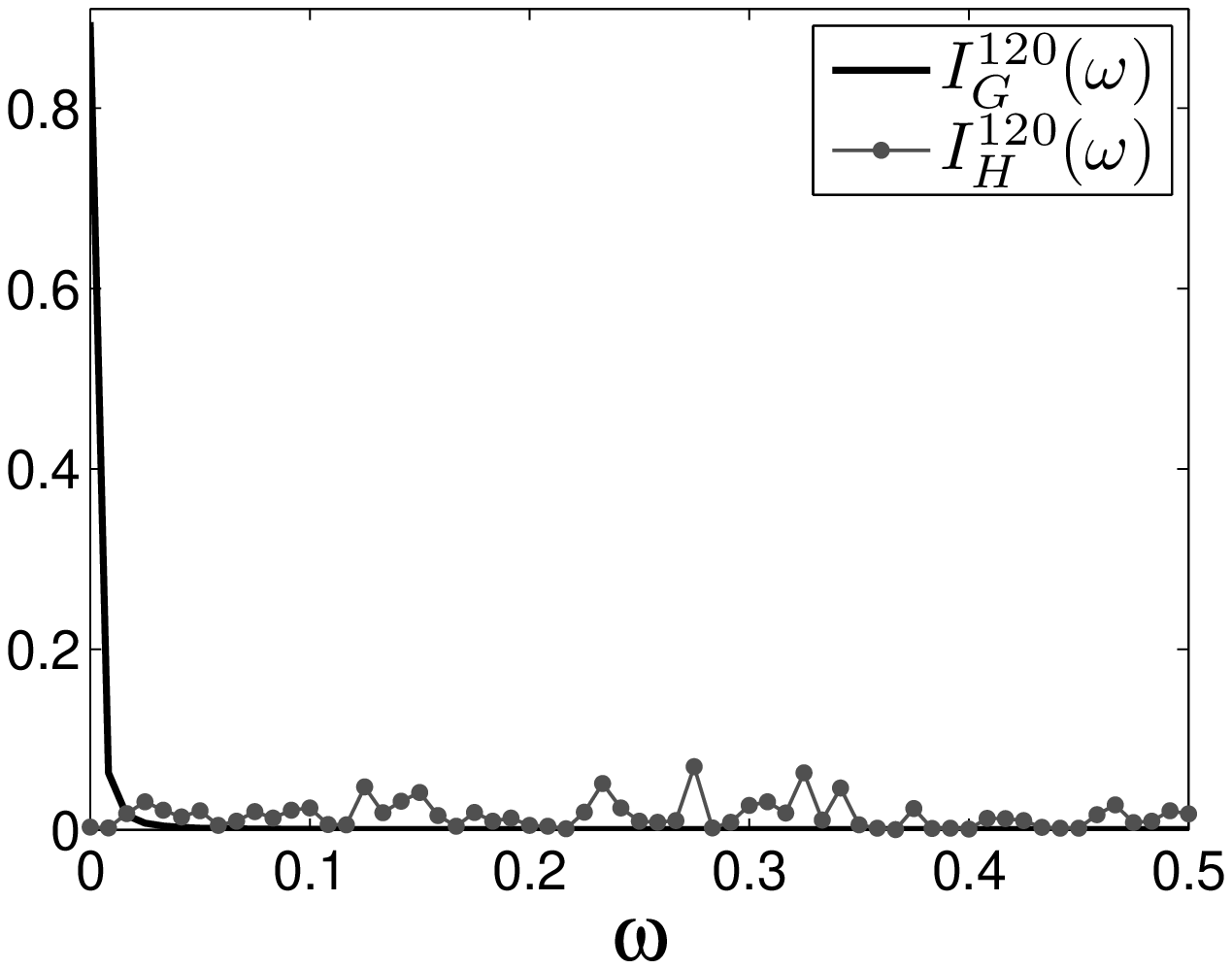}}
	\end{minipage}
\end{minipage}\\

\vspace{10pt}

\begin{minipage}{1.1\textwidth}
	\begin{minipage}{0.3\textwidth}
	\centerline{\textsf{\small {\bf d)} Periodograms of}}
	\centerline{\textsf{\small $G$ and $H$ (zoomed)}}
	\centerline{\includegraphics[width=0.99\textwidth]{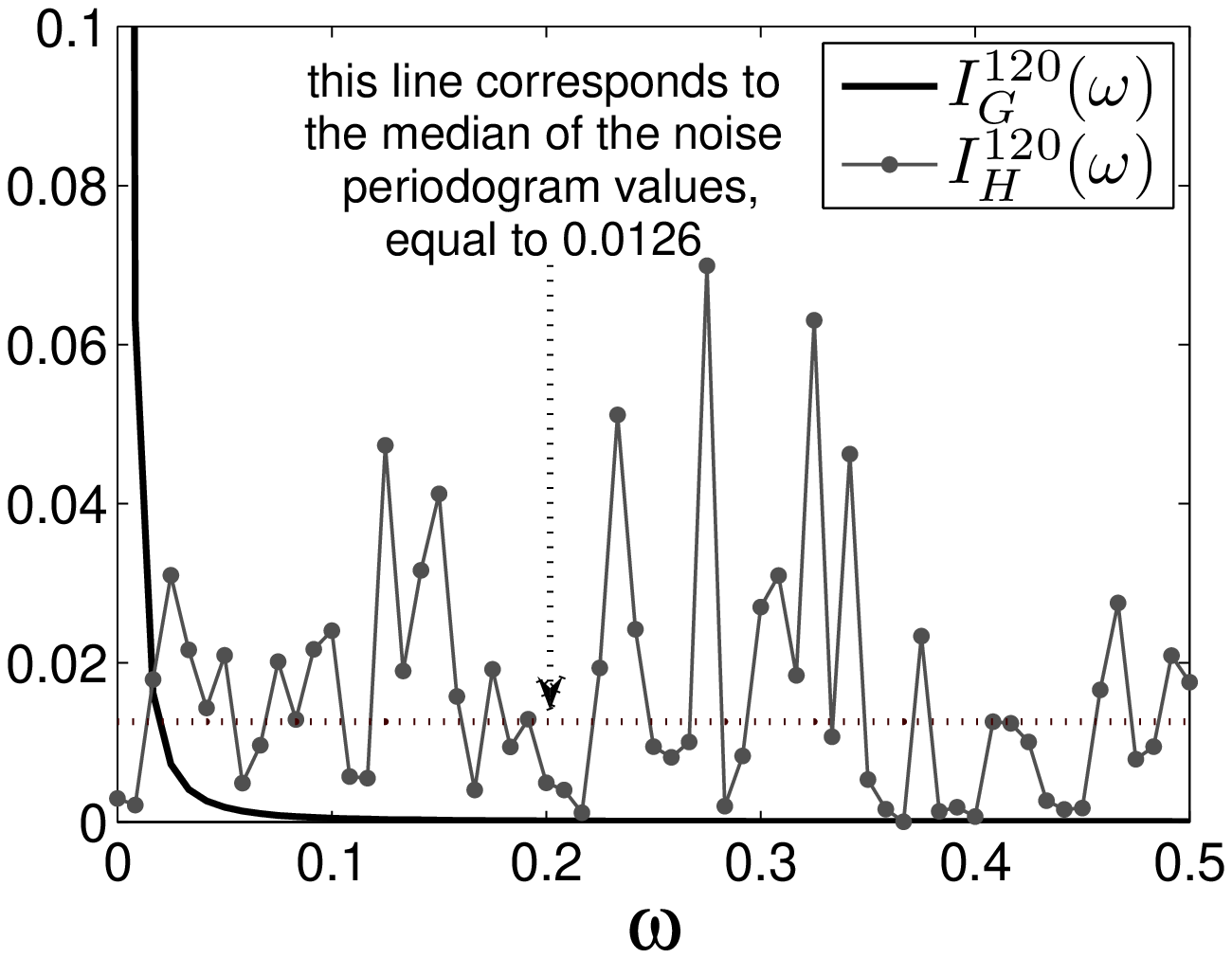}}
	\end{minipage}
	\begin{minipage}{0.3\textwidth}
	\centerline{\textsf{\small {\bf e)} Periodogram of $F$ and}}
	\centerline{\textsf{\small its estimates}}
	\centerline{\includegraphics[width=0.99\textwidth]{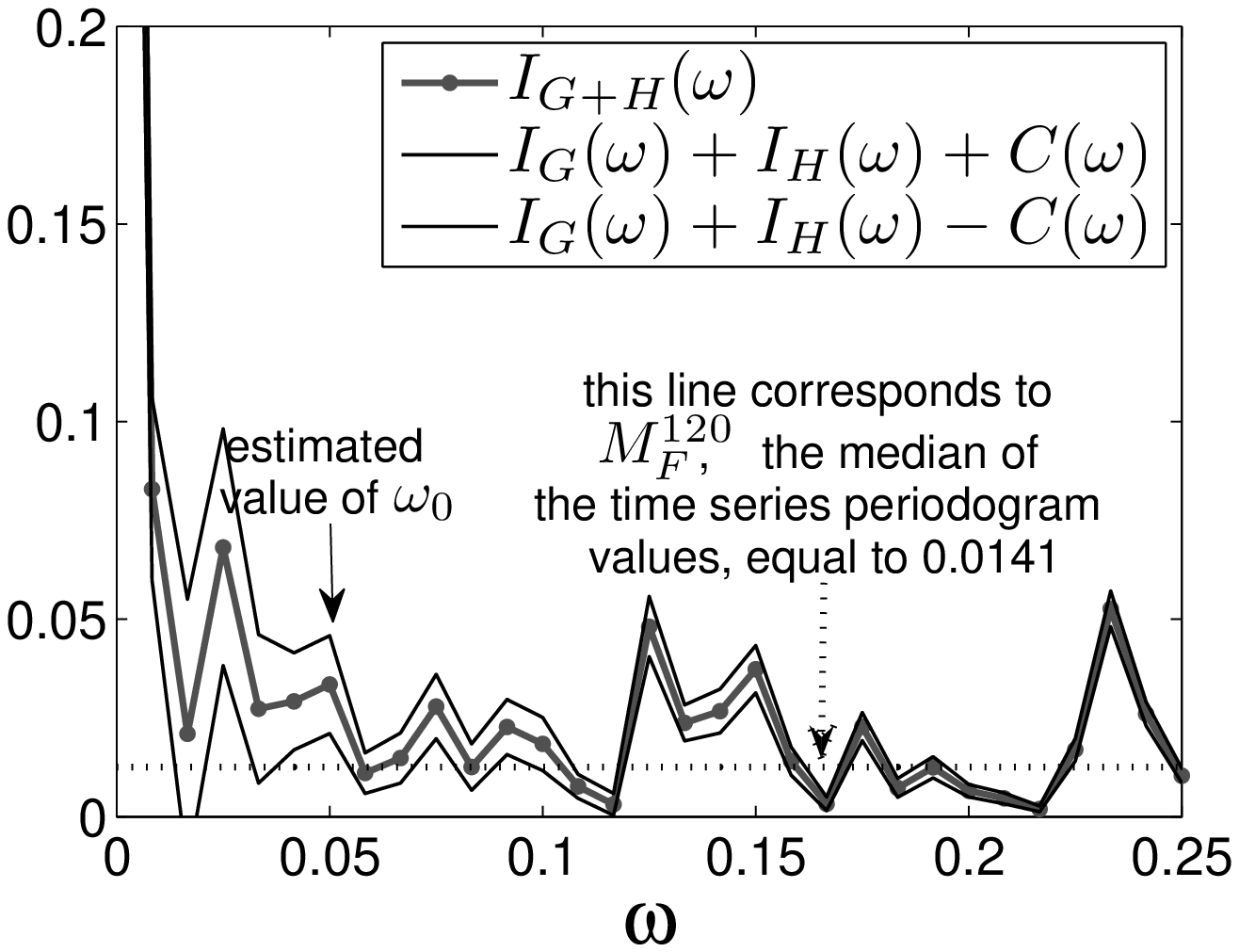}}
	\end{minipage}
	\begin{minipage}{0.3\textwidth}
	\centerline{}
	\centerline{\textsf{\small {\bf f)} Discrepancy}}
	\centerline{\includegraphics[width=0.99\textwidth]{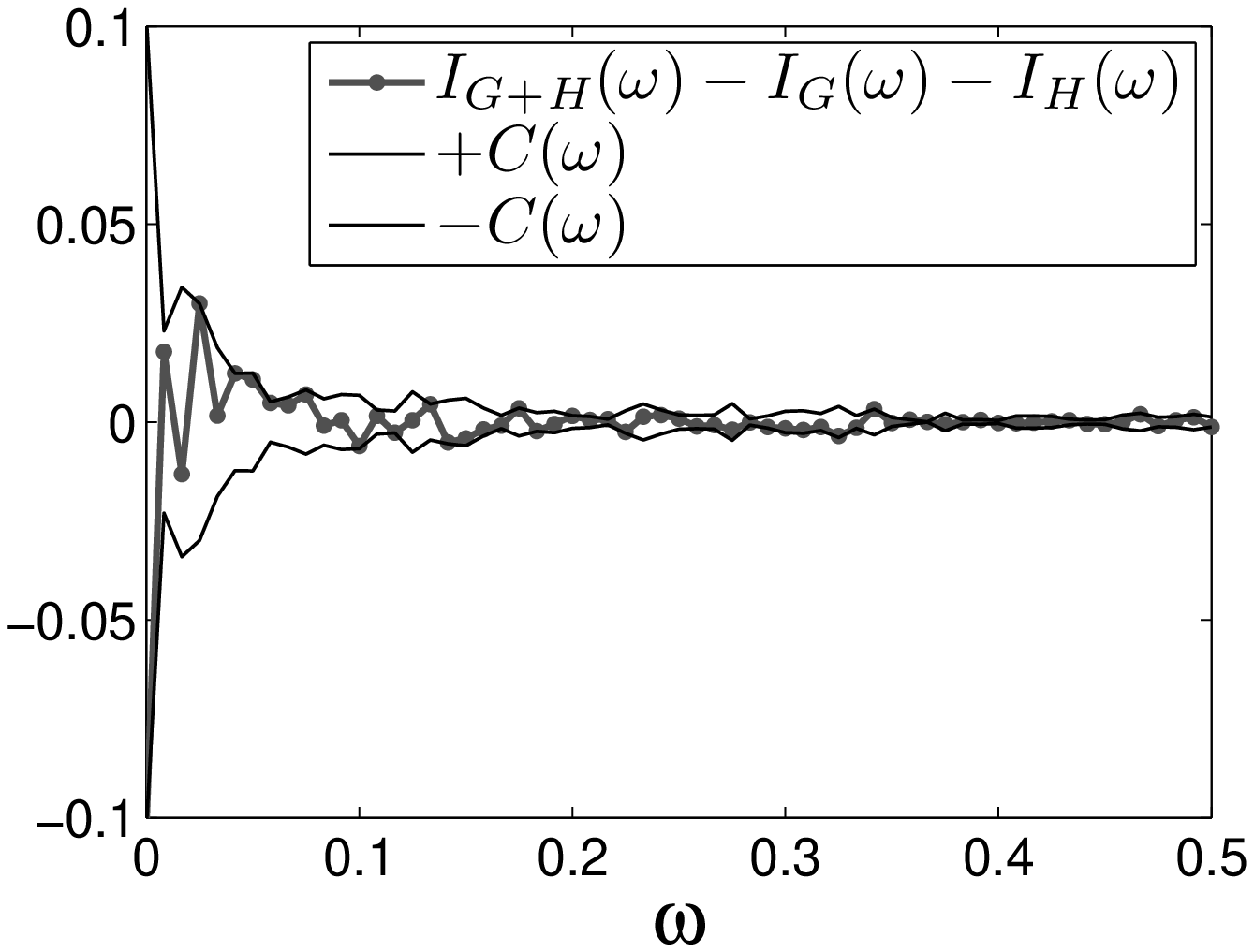}}
	\end{minipage}
\end{minipage}
\end{center}
\caption{The choice of $\omega_0$ for an exponential trend and Gaussian noise;
The value $C(\omega)$ used in the legends is equal to $2\sqrt{\PGN(\omega)\PHN(\omega)}$.\label{fig:LF.Pi(G+H)}}
\end{figure}
\vspace*{5mm}

\end{example}

\section{THE LOW-FREQUENCY CONTRIBUTION $\mathcal{C}_0$}
\label{sec:param_C0}
Before suggesting a procedure for selection of the second parameter of the proposed method, 
the low-frequency threshold $\Cnl$,
we investigate the effect of the choice of $\Cnl$ on the quality of the trend extracted. For this aim, 
we consider a time series model with a trend that generates SVD components with known numbers. 
Then, for a sufficient number of simulated time series, we compare our trend extraction
procedure with a SSA-based procedure which simply reconstructs the trend using the known 
trend SVD components. 

\subsection{A simulation example: an exponential trend plus a Gaussian noise}
\label{sec:C0_exptrendnoise}
The model considered is the same as in example above. Let the time series $F=(f_0,\ldots,f_{N-1})$
consist of an exponential trend $t_n$ plus a Gaussian white noise $r_n$:
\begin{equation}
\label{eq:C0_simmodel}
f_n=t_n + r_n, \quad t_n=e^{\alpha n},\quad r_n=\sigma e^{\alpha n} \varepsilon_n, ~ \varepsilon_n \sim iidN(0,1).
\end{equation}
According to~\cite{GNZ01}, for such a time series with moderate noise the first SVD component corresponds to the trend.
We considered only the noise levels when this is true (empirically checked). Note that the noise $r_n$ 
has a multiplicative model as its standard deviation is proportional to the trend. 

In the following, we consider the following properties. 
First, we calculate the difference between the trend
$\hat t_n(\Cnl)$ resulted from our method with $\Cnl$ used and the reconstruction 
$\tilde t_n$ of the first SVD component exploiting 
the weighted mean square error (MSE) because this measure is more relevant for a model with a multiplicative noise
than a simple MSE:
\begin{equation}
	\mathcal{D}(\hat t_n(\Cnl), \tilde t_n) = \frac{1}{N} \sum_{n=0}^{N-1} e^{-2\alpha n} \bigl( \hat t_n(\Cnl) - \tilde t_n \bigr)^2.
	\label{eq:C0_D1}
\end{equation}
This measure compares our trend and the ideal SSA trend. 
Second, we calculate the weighted mean square errors between $\hat t_n(\Cnl)$, $\tilde t_n$ and the true trend $t_n$ separately:
\begin{equation}
	\mathcal{D}(\hat t_n(\Cnl)) = \frac{1}{N} \sum_{n=0}^{N-1} e^{-2\alpha n} \bigl( t_n - \hat t_n(\Cnl) \bigr)^2, \quad
	\mathcal{D}(\tilde t_n) = \frac{1}{N} \sum_{n=0}^{N-1} e^{-2\alpha n} \bigl( t_n - \tilde t_n \bigr)^2.
	\label{eq:C0_D12}
\end{equation}

\subsubsection{Scheme of estimation of the errors using simulation}
The errors (\ref{eq:C0_D1}), (\ref{eq:C0_D12}) are estimated using the following scheme. 
We simulate $S$ realizations of the time series $F$ according to the model~(\ref{eq:C0_simmodel})
and calculate the mean of $\D(\hat t_n(\Cnl), \tilde t_n)$ for all values of $\Cnl$ from the large grid $0{:}0.01{:}1$:
\begin{equation}
	\overline{\D}(\hat t_n(\Cnl), \tilde t_n) = \frac{1}{S} \sum_{s=1}^S \D(\hat t_n^{(s)}(\Cnl), \tilde t^{(s)}_n),
	\label{eq:C0_meanD}
\end{equation}
where $\hat t_n^{(s)}(\Cnl)$ and $\tilde t_n^{(s)}$ denote trends of the $s$-th simulated time series. 
The mean errors $\overline{\D}(\hat t_n(\Cnl))$, $\overline{\D}(\tilde t_n)$ between 
the true trend $t_n$ and the extracted trends $\hat t_n(\Cnl)$ and $\tilde t_n$, respectively, are calculated similarly. 
Let us also denote the minimal values of the mean errors as
\begin{equation}
	\overline{\D}^{min}(\hat t_n, \tilde t_n) = \min_{\Cnl} \overline{\D}(\hat t_n(\Cnl), \tilde t_n), \quad
	\overline{\D}^{min}(\hat t_n) = \min_{\Cnl} \overline{\D}(\hat t_n(\Cnl))
	\label{eq:C0_minmeanD}
\end{equation}
and the value of $\Cnl$ providing the minimal mean error between the extracted trend and the ideal SSA trend as
\begin{equation*}
	\Cnl^{opt} = \arg \min_{\Cnl} \overline{\D}(\hat t_n(\Cnl), \tilde t_n),
\end{equation*}
so that $\overline{\D}^{min}(\hat t_n, \tilde t_n)=\overline{\D}(\hat t_n(\Cnl^{opt}), \tilde t_n)$.

The simulated time series are of length $N=47$. In order to achieve the best separability~\cite{GNZ01}
we have selected the SSA window length $L=\lceil N/2 \rceil=24$. The estimates of the mean errors are
calculated on $S=10^4$ realizations of the time series.

We consider different values of the model parameters $\alpha$ and $\sigma$.
The values of $\alpha$ are 0 (corresponding to a constant trend), 0.01 and 0.02 which correspond to the increase of trend values (from $t_0$ 
to $t_{N-1}$) in 1, 1.6 and 2.5 times, respectively. The levels of noise are $0.2 \leqslant \sigma \leqslant 1.6$. 
It was empirically checked that for such levels of noise the first SVD component corresponds to the trend.

Moreover, we estimated the probability of the type I error of not selecting the first SVD component as 
the ratio of times when the first component is not identified as a trend component by our procedure
to the number of repetitions $S$.

\begin{figure}[ht]
\centerline{\includegraphics[width=0.55\textwidth]{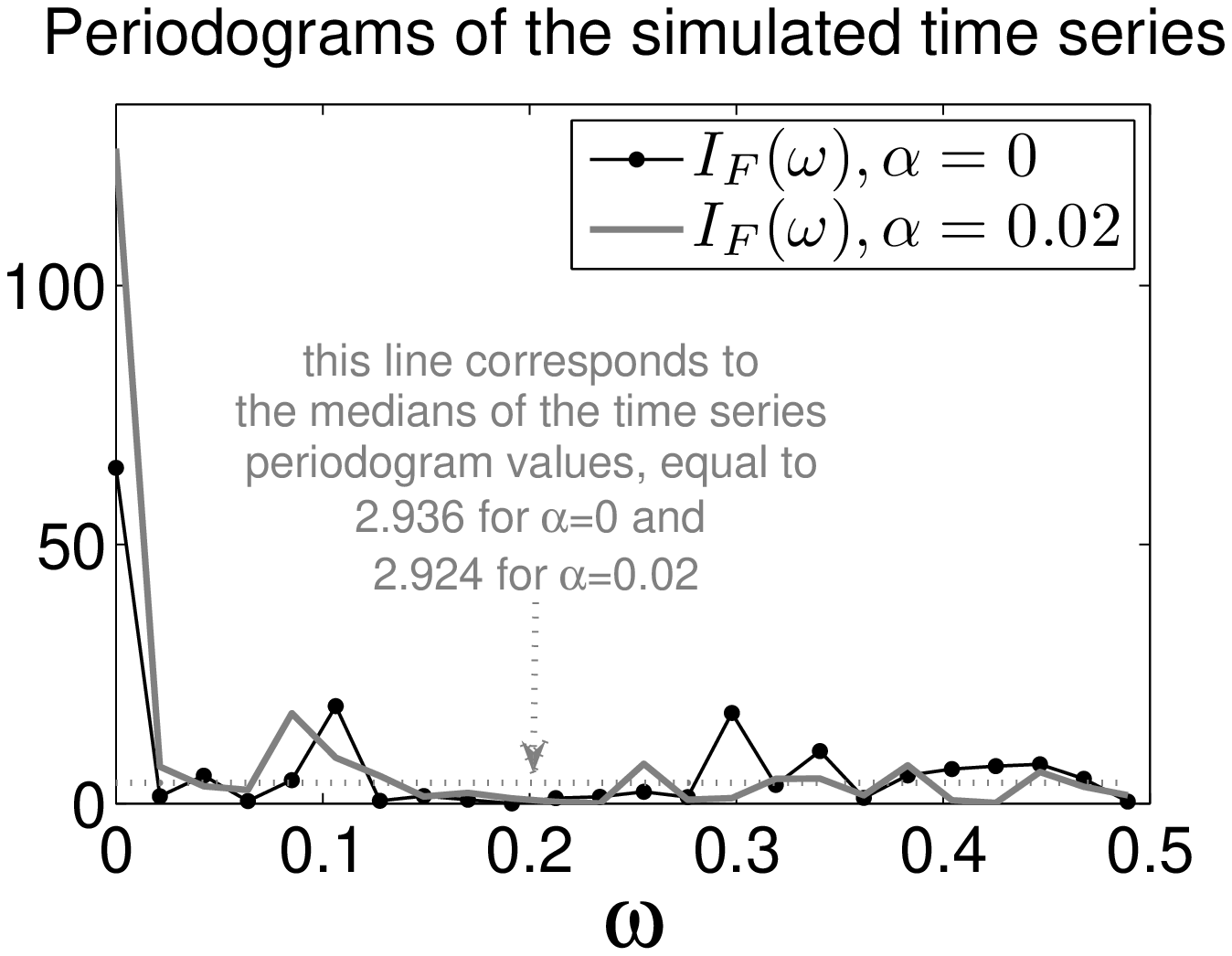}}	
	\caption{The periodograms of two time series of the model~(\ref{eq:C0_simmodel}) with $\sigma=1.6$ and       
	$\alpha=0,0.02$.\label{fig:C0_periodogrsimulated}}
\end{figure}

\paragraph{Choice of $\omega_0$.}
In order to select the low-frequency threshold $\omega_0$, we considered several simulated time series with 
different $\alpha$ and the maximal noise $\sigma=1.6$. Two examples of their periodograms for $\alpha=0$ and 
$\alpha=0.02$ are depicted in Figure~\ref{fig:C0_periodogrsimulated}.
The median values for the periodograms depicted in Figure~\ref{fig:C0_periodogrsimulated} are 2.936 and 2.924 
which leads to $\omega_0=0$ for $\alpha=0$ and $\omega_0=\lceil 1/N \cdot L \rceil / L = 1/24 \approxeq 0.042$ for $\alpha=0.02$
estimated using (\ref{eq:w0_LN}).
We decided to take the same $\omega_0=0.042$ (the largest one) for all $\alpha$ considered.

\subsubsection{Simulation results}
\begin{figure}[ht]
\centerline{\includegraphics[width=0.48\textwidth]{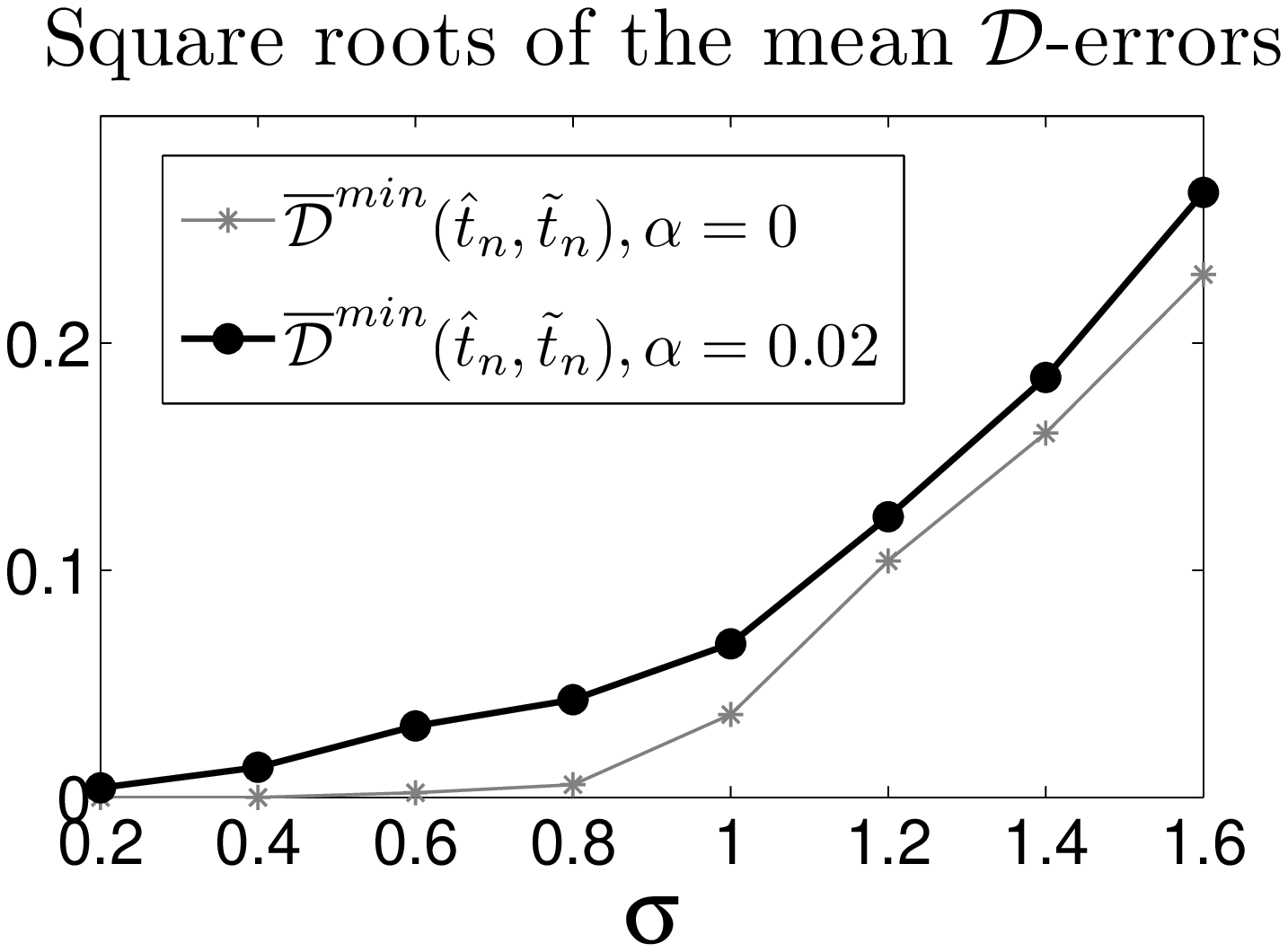} \quad
		\includegraphics[width=0.48\textwidth]{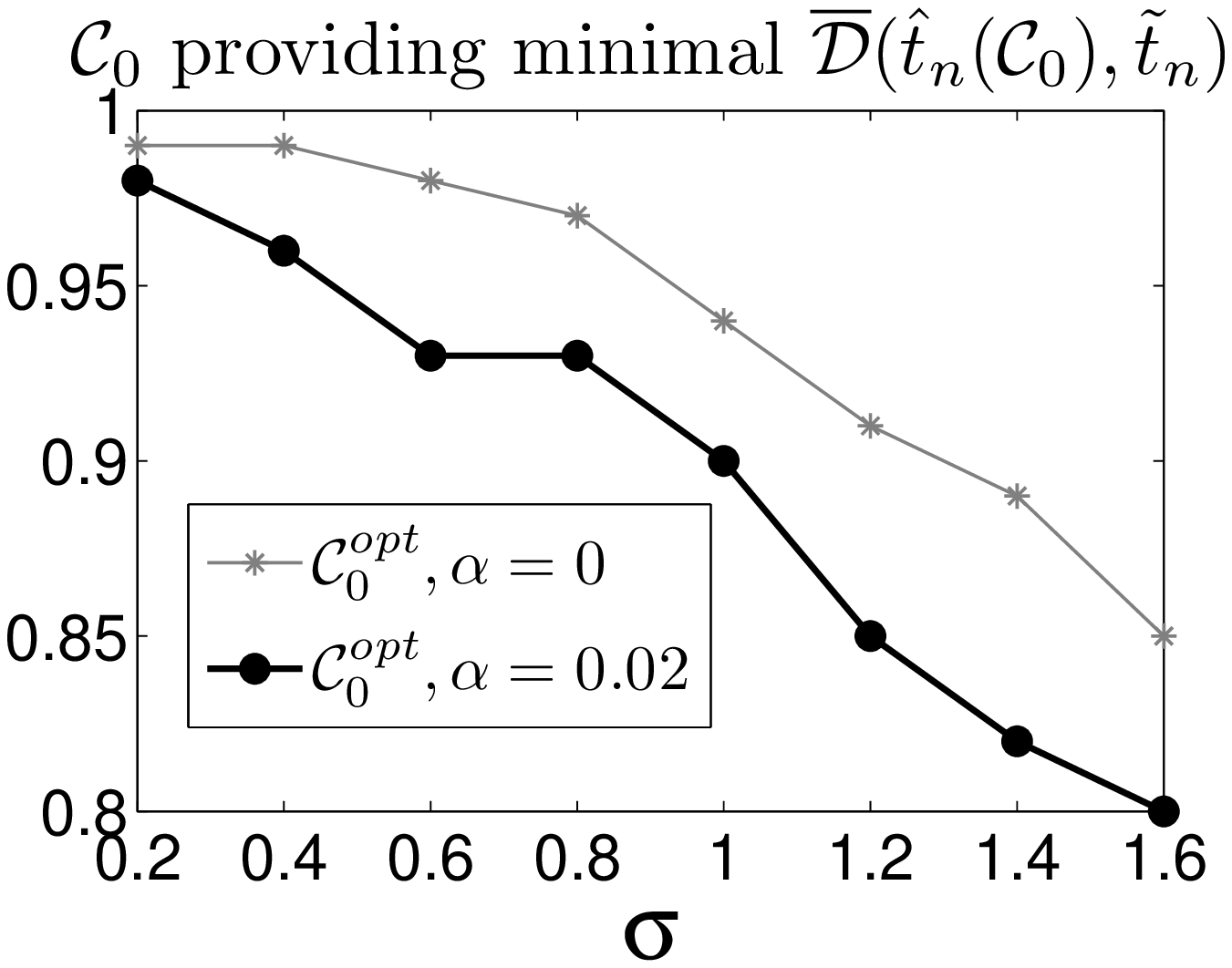}}
\centerline{\includegraphics[width=0.48\textwidth]{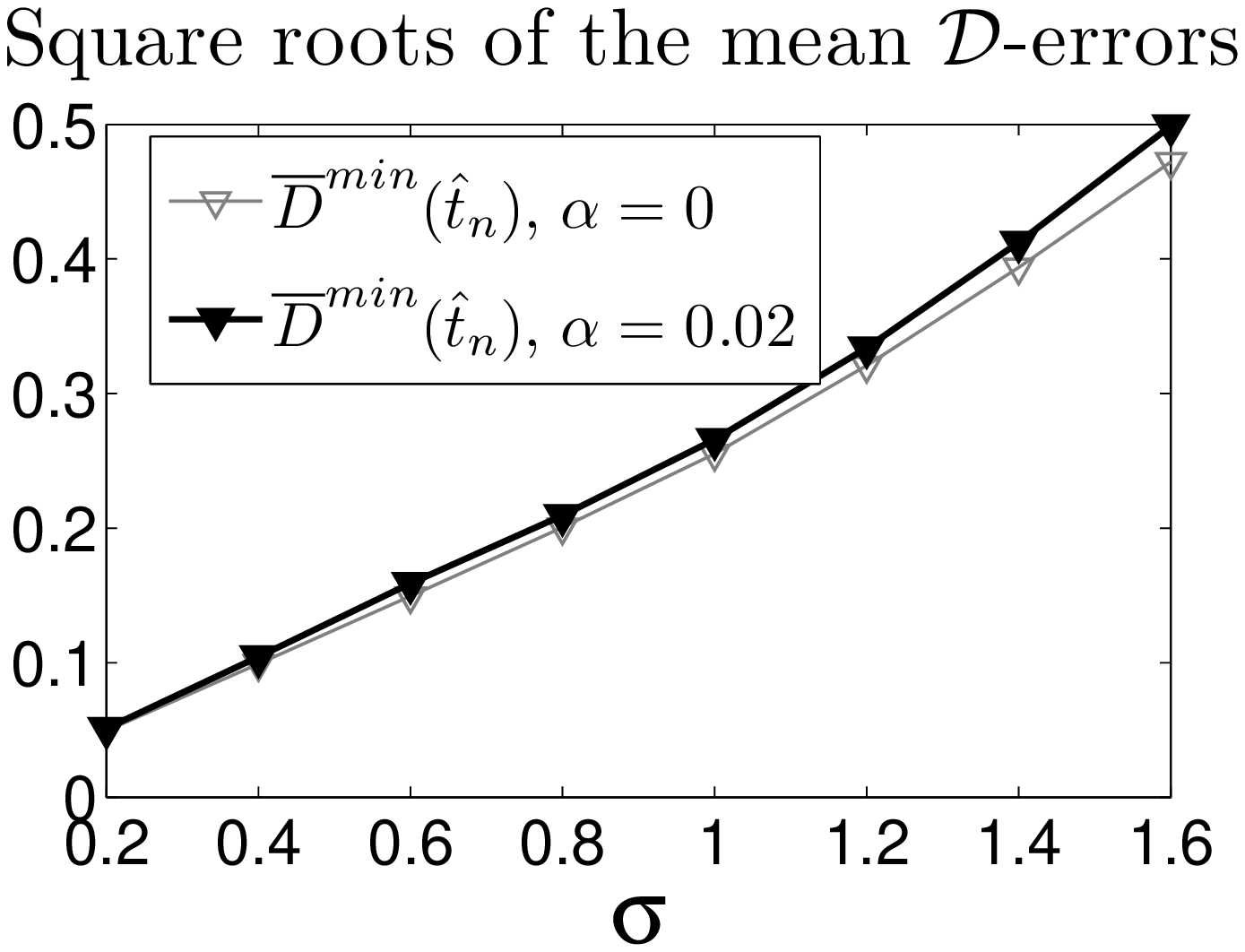} \quad
    \includegraphics[width=0.48\textwidth]{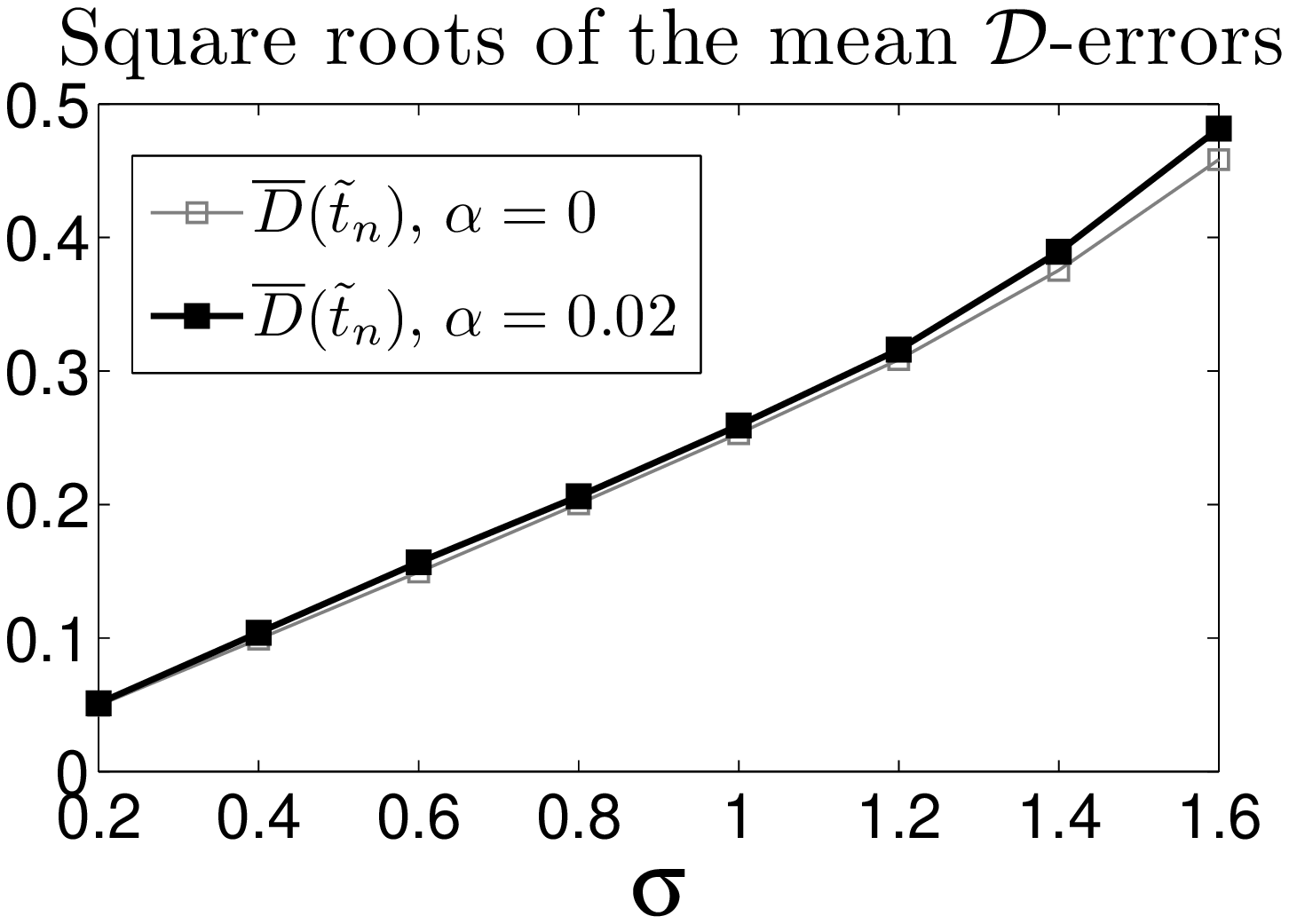}}	
	\caption{The square roots of the mean errors $\overline{\D}^{min}(\hat t_n, \tilde t_n)$ (top left)
	$\overline{\D}^{min}(\hat t_n)$ (bottom left) and	$\overline{\D}(\tilde t_n)$ (bottom right) 
	as well as the optimal $\Cnl$ value providing a minimal mean error $\overline{\D}^{min}(\hat t_n, \tilde t_n)$ 
	between the extracted trend	and the ideal SSA trend (top right); all for $\alpha=0$ and $\alpha=0.02$. \label{fig:C0_Derrors_al0al002}}
\end{figure}

Figure~\ref{fig:C0_Derrors_al0al002} shows the evolution of the square roots of the mean errors 
and $\Cnl^{opt}$ as a function of $\sigma$.
The values $\alpha=0$ and $\alpha=0.02$ are used.
The square roots of the mean errors (i.e. standard deviations) are taken for better comparison with $\sigma$ 
which is the standard deviation multiplier of the noise.

The plots of the minimal mean error $\overline{\D}^{min}(\hat t_n, \tilde t_n)$ and the optimal $\Cnl^{opt}$ for $\alpha=0.02$ 
are depicted in Figure~\ref{fig:C0_Derrors_al0al002}, where the values for $\alpha=0$ are also shown in gray color. The estimates
for $\alpha=0.01$ are not reported here.

The interpretation of the produced results is as follows. First, the trend extracted 
with the optimal $\Cnl$ is very similar to the ideal SSA trend, reconstructed by 
the first SVD component since $\overline{\D}^{min}(\hat t_n, \tilde t_n) \ll \overline{\D}^{min}(\hat t_n)$
(the error between our trend and the ideal trend is much smaller than the error of the ideal trend itself),
especially when $\sigma\leqslant 0.8$.
Moreover, the estimated probability of the type I error (i.e. the probability of not selecting the first SVD component) is 
less than 0.05 for $\sigma\leqslant 1.4$. All this allows us to conclude that in case of an exponential
trend and a white Gaussian noise the proposed method of trend extraction with an optimal $\Cnl$ 
with high probability selects the required first SVD component corresponding to the trend.

The trend $\hat t_n(\Cnl^{opt})$ extracted with an optimal $\Cnl$ estimates the true
trend quite good when comparing the deviation $\sqrt{\overline{\D}^{min}(\hat t_n)}$ with 
the noise standard deviation $\sigma$. For example, for $\sigma=1.6$ the value 
of $\sqrt{\overline{\D}^{min}(\hat t_n)}$ is approximately equal to $0.5$.

Note that for different $\alpha$ the mean errors $\overline{\D}^{min}(\hat t_n)$
are very similar though the used optimal values of $\Cnl$ are quite different (Figure~\ref{fig:C0_Derrors_al0al002}). 
This shows that the method adapts to the change of the model parameter $\alpha$.

\begin{figure}[ht]
\centerline{
\includegraphics[width=0.55\textwidth]{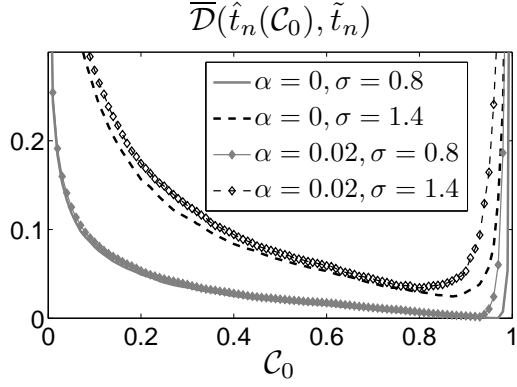}}
	\caption{The error $\overline{\D}(\hat t_n(\Cnl), \tilde t_n)$ as a function of $\Cnl$
	for different combinations of $\alpha=0,0.02$ and $\sigma=0.8,1.4$. \label{fig:Derror_C0}}
\end{figure}

Let us consider the dependence of inaccuracy of the proposed trend extraction method
on the value of $\Cnl$. As above, the inaccuracy is measured with the minimal 
mean error $\overline{\D}^{min}(\hat t_n, \tilde t_n)$
between the extracted trend and the ideal SSA trend. Figure~\ref{fig:Derror_C0} shows
the graphs of this error as a function of $\Cnl$ for different exponentials $\alpha$ 
and noise levels $\sigma$.
\vspace*{5mm}
\vspace*{5mm}

One can see that it is crucial not to select too large $\Cnl$ since in this case the trend component
can be not included in the reconstruction (that is also confirmed by the estimated probability of the type I error 
which is not reported here). At the same time without significant loss of accuracy one can 
choose $\Cnl$ smaller than $\Cnl^{opt}$ (corresponding to the best accuracy). This is true due to 
the small contribution of each of noise components which can be erroneously included for $\Cnl<\Cnl^{opt}$.

\subsection{Heuristic procedure for the choice of $\Cnl$}
Based on the observations of section~\ref{sec:C0_exptrendnoise}, we propose
the following heuristic procedure for choosing the value of the method low-frequency 
threshold $\mathcal{C}_0$.

As discussed, trend EOFs vary slow. First we show that this property is inherited 
by the trend elementary reconstructed components, the time series components each 
reconstructed from one trend SVD component.
\begin{proposition}
\label{prop:LF.LFincome_reconstrts}
Let $(\sqrt{\lambda}, U, V)$ be an eigentriple of SSA decomposition of a time series $F$,
$U=(u_1,\ldots,u_L)\transp$, $V=(v_1,\ldots,v_L)\transp$, and $\tldFone$ be a time series 
reconstructed by this eigentriple. If it is true that
\begin{equation*}
\exists \delta_1, \delta_2 \in \mathbb{R}: ~~ \forall k, ~ 1\leqslant k \leqslant L-1:
\quad
|u_{k+1}-u_{k}|<\delta_1, ~ |v_{k+1}-v_{k}|<\delta_2,
\end{equation*}
then for the elements of $\tldFone=(\tldfone_0,\ldots,\tldfone_{N-1})$ the following holds:
\begin{equation*}
\exists \epsilon(\delta_1, \delta_2): ~~ \forall n, ~ L^*-1\leqslant n < K^*:
\quad
\Bigl|\tldfone_{n+1}-\tldfone_n\Bigr|<\epsilon(\delta_1, \delta_2),
\end{equation*}
where ${L}^* = \min\{L,K\}$, ${K}^* = \max\{L,K\}$.
\end{proposition}
\begin{proof}[Proof of Proposition~\ref{prop:LF.LFincome_reconstrts}]
One can easily prove this proposition taking into account how the elementary
reconstructed component $\tldFone$ is constructed from its eigentriple $(\sqrt{\lambda}, U, V)$,
see section~\ref{sec:SSA}.
First, the matrix ${\bf Y}=\sqrt{\lambda} U V\transp$ is constructed. 
Second, the hankelization of ${\bf Y}$ is performed.

Let us show how to calculate $\epsilon$ using (\ref{eq:ssa.diagaver}) for $\delta_1$, $\delta_2$ when $L\leqslant K$.
For other cases $\epsilon(\delta_1, \delta_2)$ is calculated similarly.
\begin{equation*}
\begin{array}{c}
\displaystyle \Bigl|\tldfone_{n+1}-\tldfone_{n}\Bigr| =
\frac{\sqrt{\lambda}}{L} \bigg\lvert \sum\limits_{m=1}^{L} \bigl( u_m v_{n-m+3} - u_m v_{n-m+2} \bigr) \bigg\rvert < \\
\vspace{-4mm}\\
\displaystyle  < \frac{\sqrt{\lambda}}{L} \sum\limits_{m=1}^{L} |u_m| |v_{n-m+3}-v_{n-m+2}| < \\
\vspace{-4mm}\\
\displaystyle  < \frac{\sqrt{\lambda}}{L} \delta_2 \sum\limits_{m=1}^{L} |u_m| <
\delta_2 \frac{\sqrt{\lambda}}{L}  \bigl( u_1 + (L-1) \delta_1 \bigr).
\end{array}
\end{equation*}
\end{proof}

Let us have a time series $F$ and denote its trend extracted with the 
method with parameters $\omega_0$, $\Cnl$ as $T(\omega_0,\mathcal{C}_0)$. 
In order to propose the procedure selecting $\Cnl$, we first define the normalized contribution 
of low-frequency oscillations in the residual $F-T(\omega_0,\mathcal{C}_0)$ as:
\begin{equation*}
\mathcal{R}_{F,\omega_0}(\mathcal{C}_0) = \mathcal{C}\bigl(F-T(\omega_0,\mathcal{C}_0), \omega_0\bigr) \mathcal{C}(F,\omega_0)^{-1},
\end{equation*}
where $\mathcal{C}$ is defined in equation (\ref{eq:C_Xomega}).

Based on Proposition~\ref{prop:LF.LFincome_reconstrts}, we expect that 
the elementary reconstructed components corresponding to a trend 
have large contribution of low frequencies. Thus, the maximal values 
of $\mathcal{C}_0$ which lead to selection of trend-corresponding SVD 
components should generate jumps of $\mathcal{R}_{F,\omega_0}(\mathcal{C}_0)$.

Exploiting this idea, we propose the following way of choosing $\mathcal{C}_0$:
\begin{equation}
\label{eq:Cnl_strategy}
\mathcal{C}_0^\mathcal{R}=\min \bigl\{\mathcal{C}_0\in[0,1]: \quad \mathcal{R}_{F,\omega_0}(\mathcal{C}_0+\Delta\mathcal{C}) - \mathcal{R}_{F,\omega_0}(\mathcal{C}_0) \geq \Delta\mathcal{R} \bigr\},
\end{equation}
where $\Delta\mathcal{C}$ is a search step and $\Delta\mathcal{R}$ is the given threshold.
On one hand, this strategy is heuristic and requires selection of $\Delta\mathcal{R}$,
but on the other hand, the simulation results and application to different time series
showed its ability to choose reasonable $\mathcal{C}_0$ in many cases.
Based on this empirical experience, we suggest using $0.05 \leq \Delta\mathcal{R} \leq 0.1$. The step $\Delta\mathcal{C}$
is to be chosen as small as possible to discriminate identifications occurring at different values of $\mathcal{C}_0$.
To reduce computational time, we commonly take $\Delta\mathcal{C} \geq 0.01$ and suggest a default value of $\Delta\mathcal{C}=0.01$.

\section{EXAMPLES}
\label{sec:example}
\begin{figure}[t]
     \begin{center}
    \includegraphics[width=.495\textwidth]{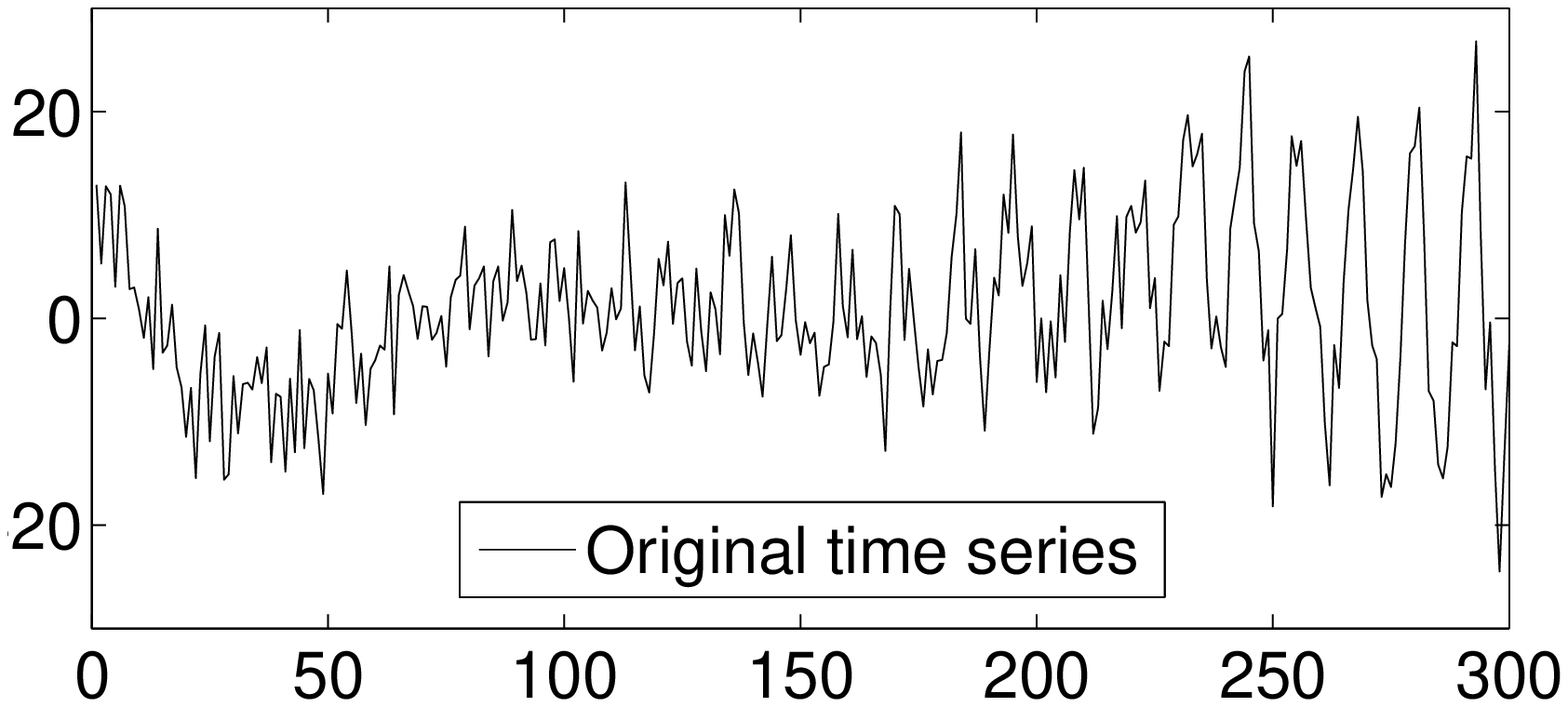}~
    \includegraphics[width=.495\textwidth]{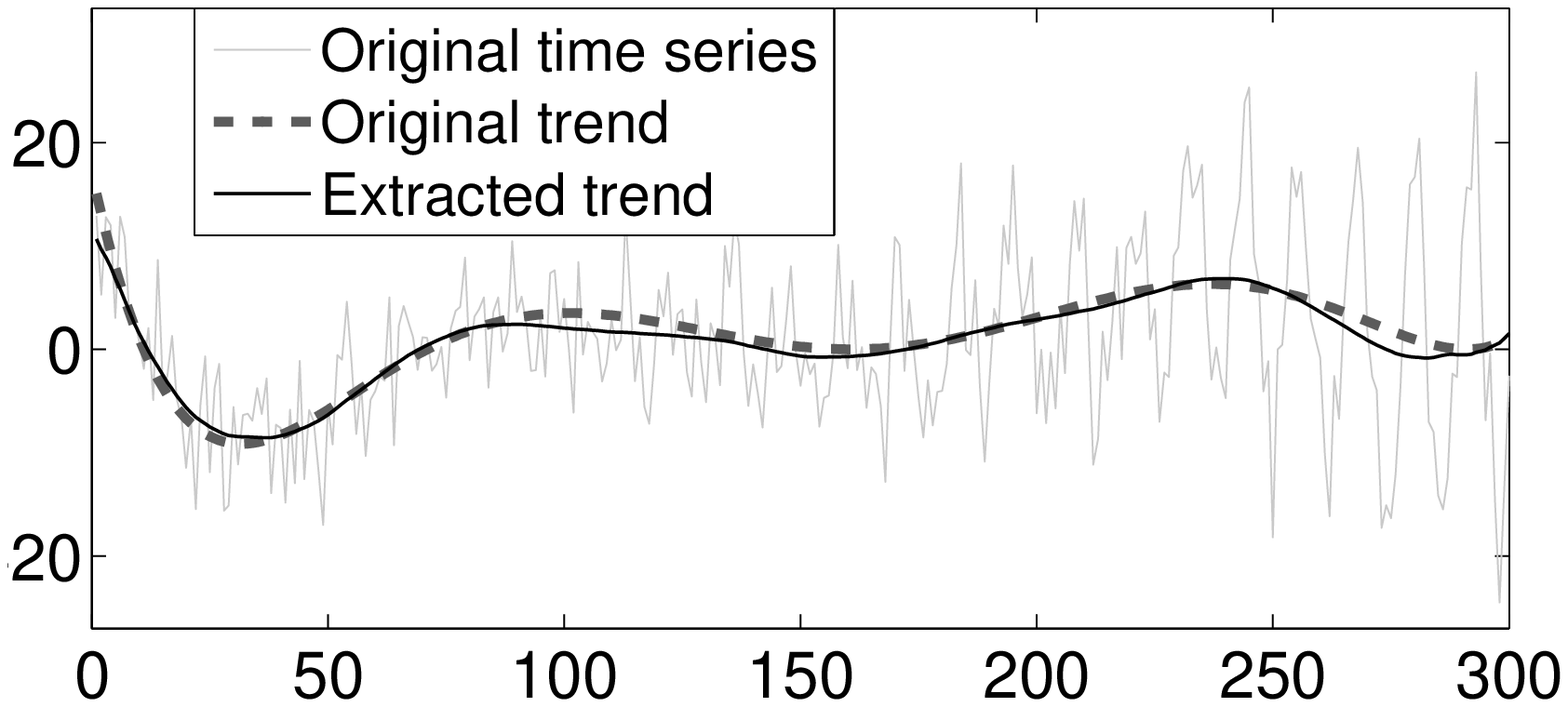}
    \includegraphics[width=.495\textwidth]{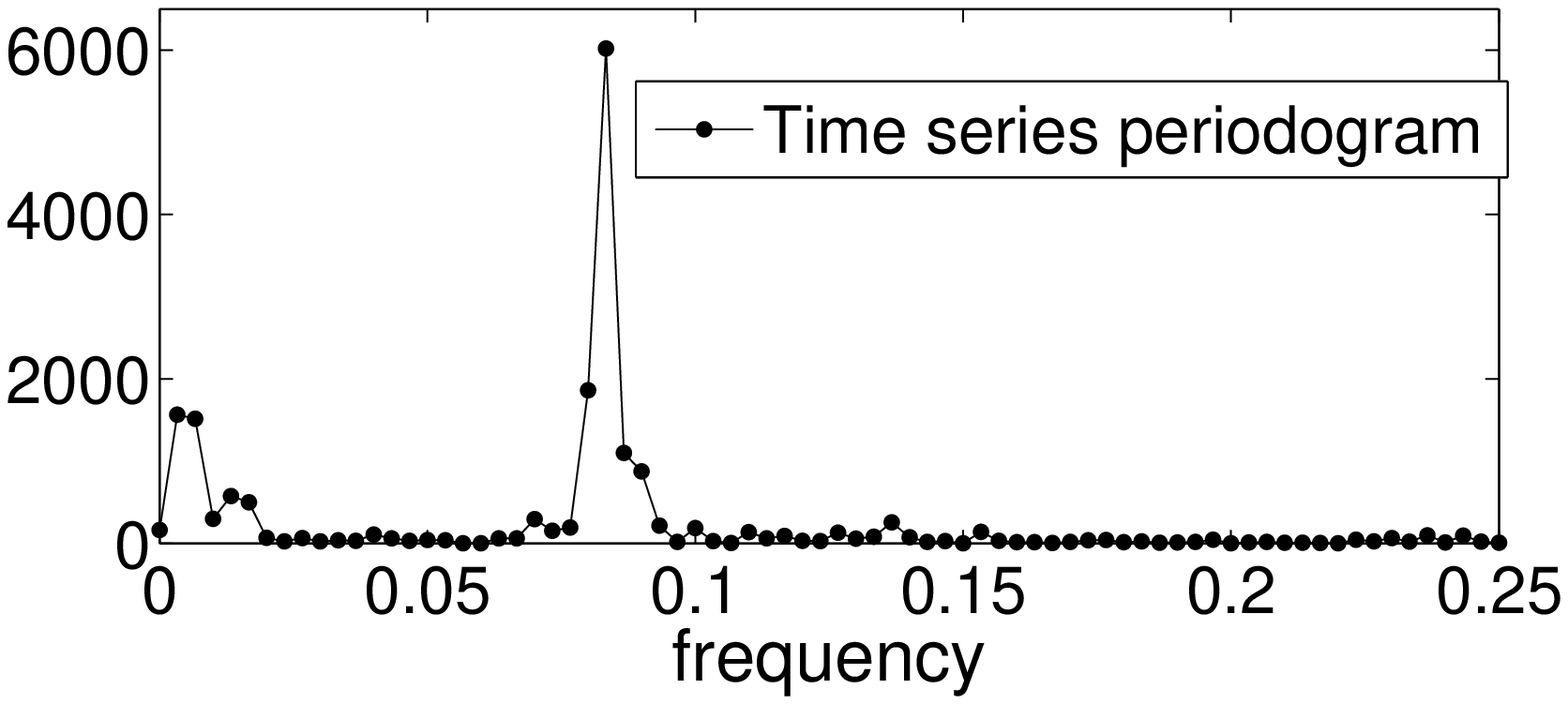}~
    \includegraphics[width=.495\textwidth]{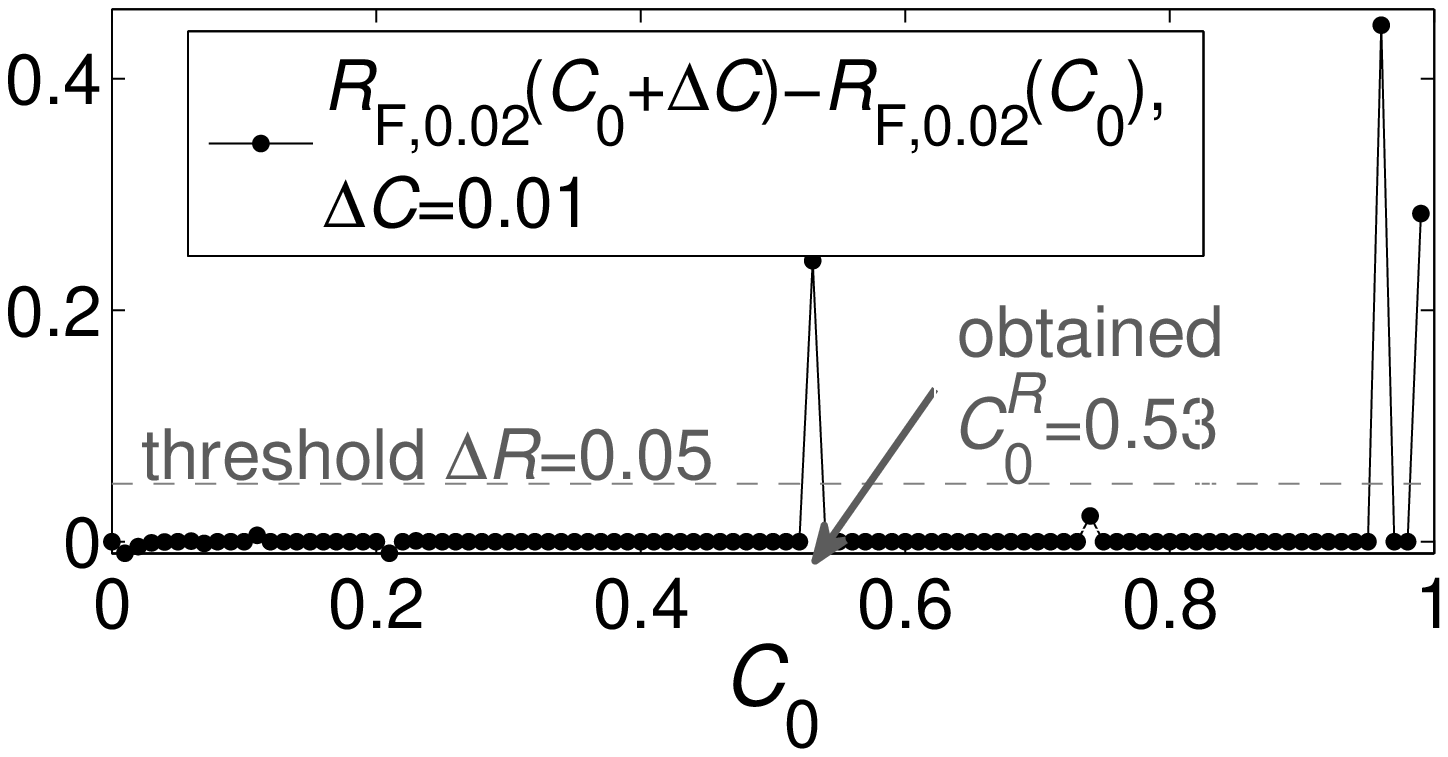}
    \end{center}
  \caption{Simulated example with a polynomial trend: original time series (top left); 
  the original trend and an extracted one with $L=180$, $\Delta\mathcal{C}=0.01$, 
  and $\Delta\mathcal{R}=0.05$ (top right); zoomed time series periodogram 
  inside $\omega\in[0,0.25]$ (bottom left); 
  the values of $\mathcal{R}_{F,\omega_0}(\mathcal{C}_0+\Delta\mathcal{C}) - \mathcal{R}_{F,\omega_0}(\mathcal{C}_0)$ used
   for the choice of $\mathcal{C}_0$ resulted in a value $\mathcal{C}_0^\mathcal{R}=0.53$ (bottom right). \label{fig:simex_ts_Pi}}
\end{figure}

\paragraph{Simulated example with polynomial trend.}
The first example illustrates the choice of parameters $\omega_0$ and $\mathcal{C}_0$.
We simulated a time series of length $N=300$, shown in Figure~\ref{fig:simex_ts_Pi},
containing a polynomial trend, an exponentially-modulated sine wave,
and a white Gaussian noise, whose $n$-th element is expressed as 
$f_n =\nolinebreak 10^{-11}(n-10)(n-70)(n-160)^2(n-290)^2 + \exp(0.01 n) \sin(2\pi n/12) + \varepsilon_n$, \quad
$\varepsilon_n$ is $iidN(0,5^2)$.
The period of the sine wave is assumed to be unknown.

We have chosen the window length $L=N/2=150$
for achieving better separability of trend and residual. The value $\omega_0=6/N=0.02$ was selected 
using~(\ref{eq:omega0_rule}), where the calculated median value is $M_X^N \approxeq 37.06$.
The search for $\mathcal{C}_0$ using (\ref{eq:Cnl_strategy}) has been done with step $\Delta\mathcal{C}=0.01$ and $\Delta\mathcal{R}=0.05$.
As shown in Figure~\ref{fig:simex_ts_Pi}, despite of the strong noise and oscillations,
the extracted trend 
approximates the original one very well. The achieved mean square error is $0.79$. For example,
the ideal low pass filter with the cutoff frequency $0.02$ produced the error of $3.14$.
This superiority is achieved mostly due to better approximation at the first and last 50 points of the time series.
All the calculations were performed using our Matlab-based software AutoSSA available at {http://www.pdmi.ras.ru/$\sim$theo/autossa}.

\vspace*{5mm}
\vspace*{5mm}

\begin{figure}[t]
\centerline{
    \includegraphics[width=.52\textwidth]{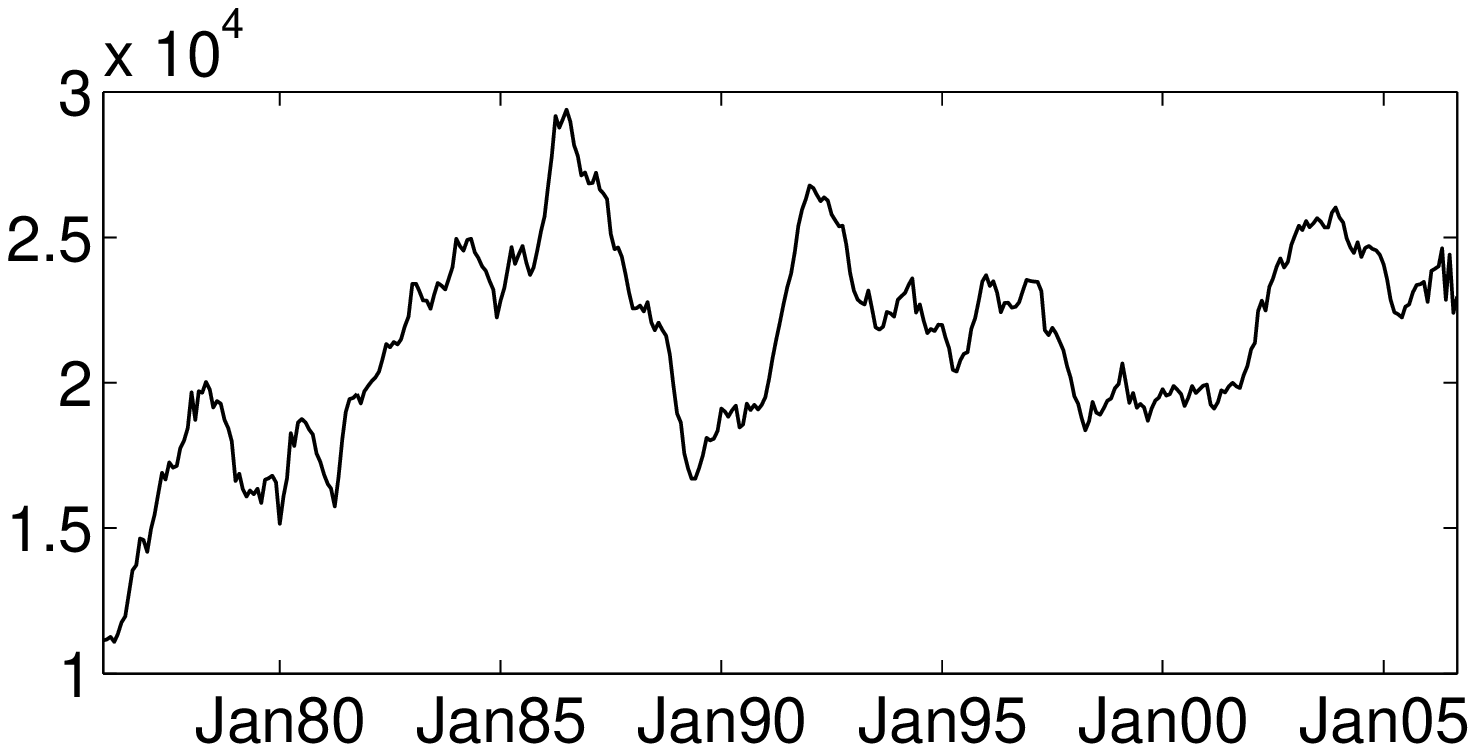}~~
    \includegraphics[width=.47\textwidth]{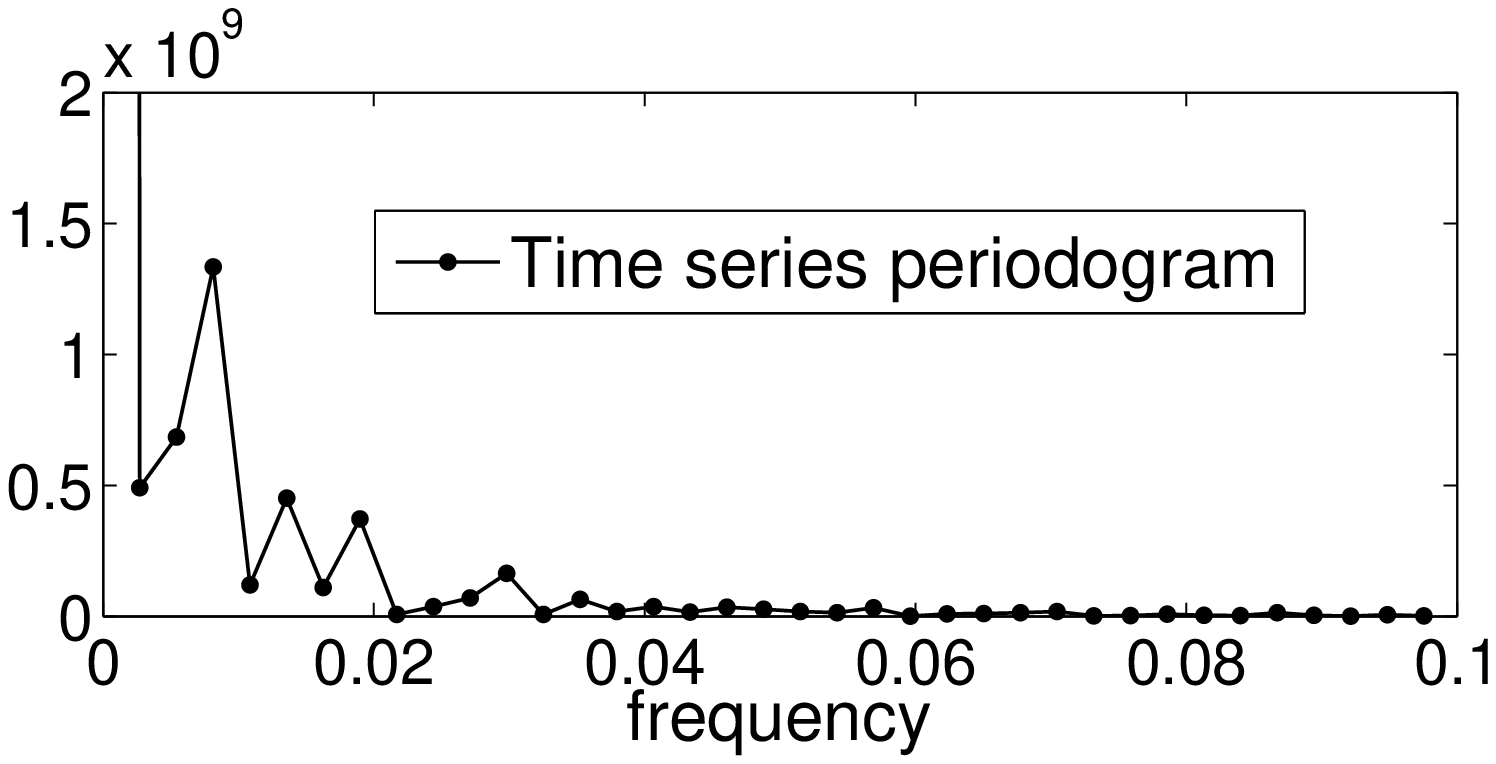}
}
  \caption{Unemployment level in Alaska: original data (left-hand side panel), zoomed periodogram (right-hand side panel).}
  \label{fig:alaska_ts}
\end{figure}

\paragraph{Trends of the unemployment level.}
Let us demonstrate extraction of trends of different scale.
We took the data of the unemployment level (unemployed persons) in Alaska for the period 1976/01-2006/09
(monthly data, seasonally adjusted), provided by the Bureau of Labor Statistics
at {http://www.bls.gov} under the identifier LASST02000004 (Figure~\ref{fig:alaska_ts}).
This time series is typical for economical applications, where data contain
relatively little noise and are subject to abrupt changes. Economists are often interested in the ``short'' term
trend which includes cyclical fluctuations and is referred to as trend-cycle.

\vspace*{5mm}
\vspace*{5mm}

The length of the data is $N=369$. For achieving better separability of trend and residual
we selected $L$ close to $N/2$ but divisible by the period $T=12$ of 
probable seasonal oscillations: $L=12 \lfloor N/24 \rfloor=180$.

\begin{figure}[t]
\begin{center}
    \includegraphics[width=.495\textwidth]{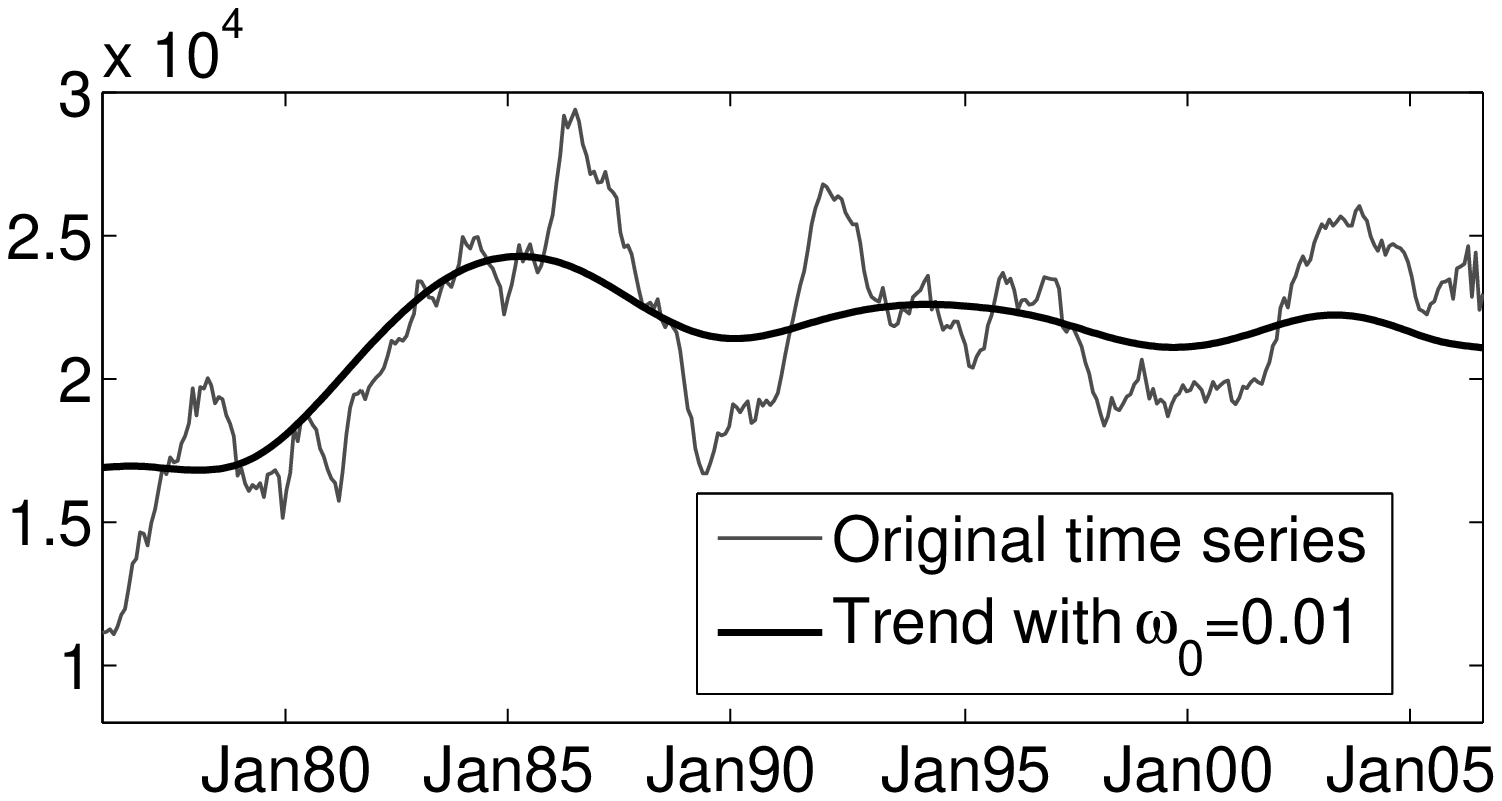}
    \includegraphics[width=.495\textwidth]{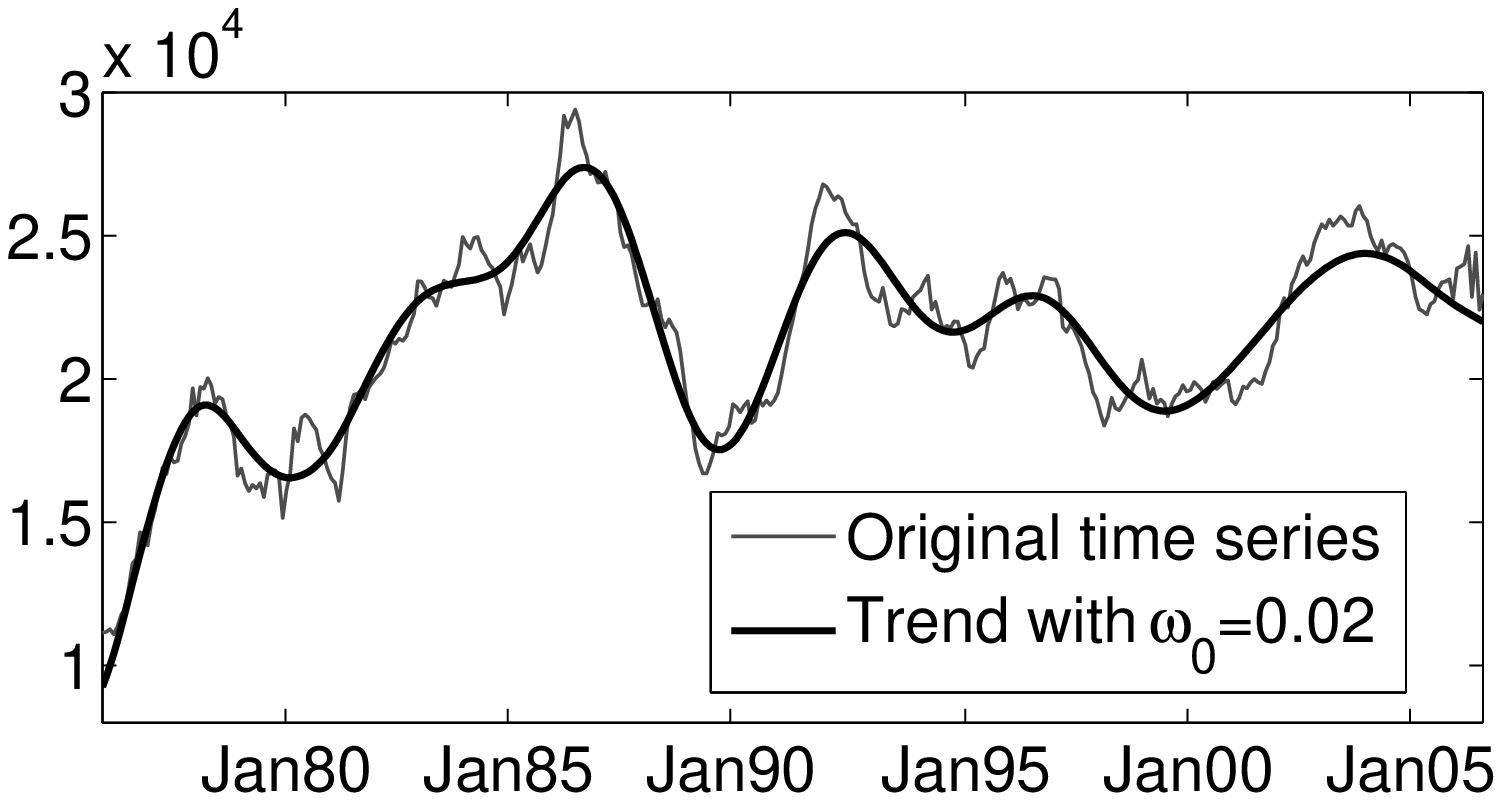} ~
    \includegraphics[width=.495\textwidth]{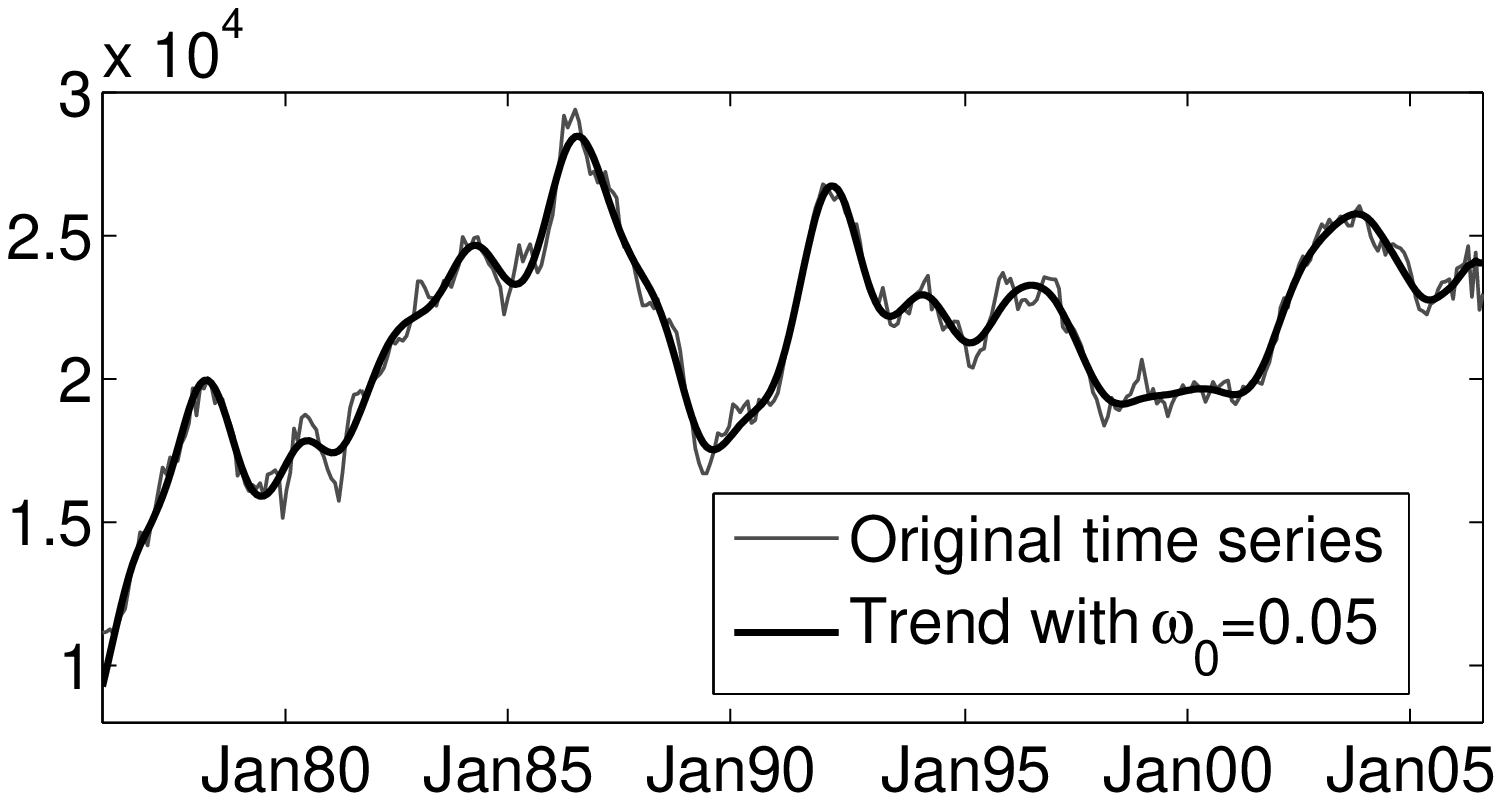}
    \includegraphics[width=.495\textwidth]{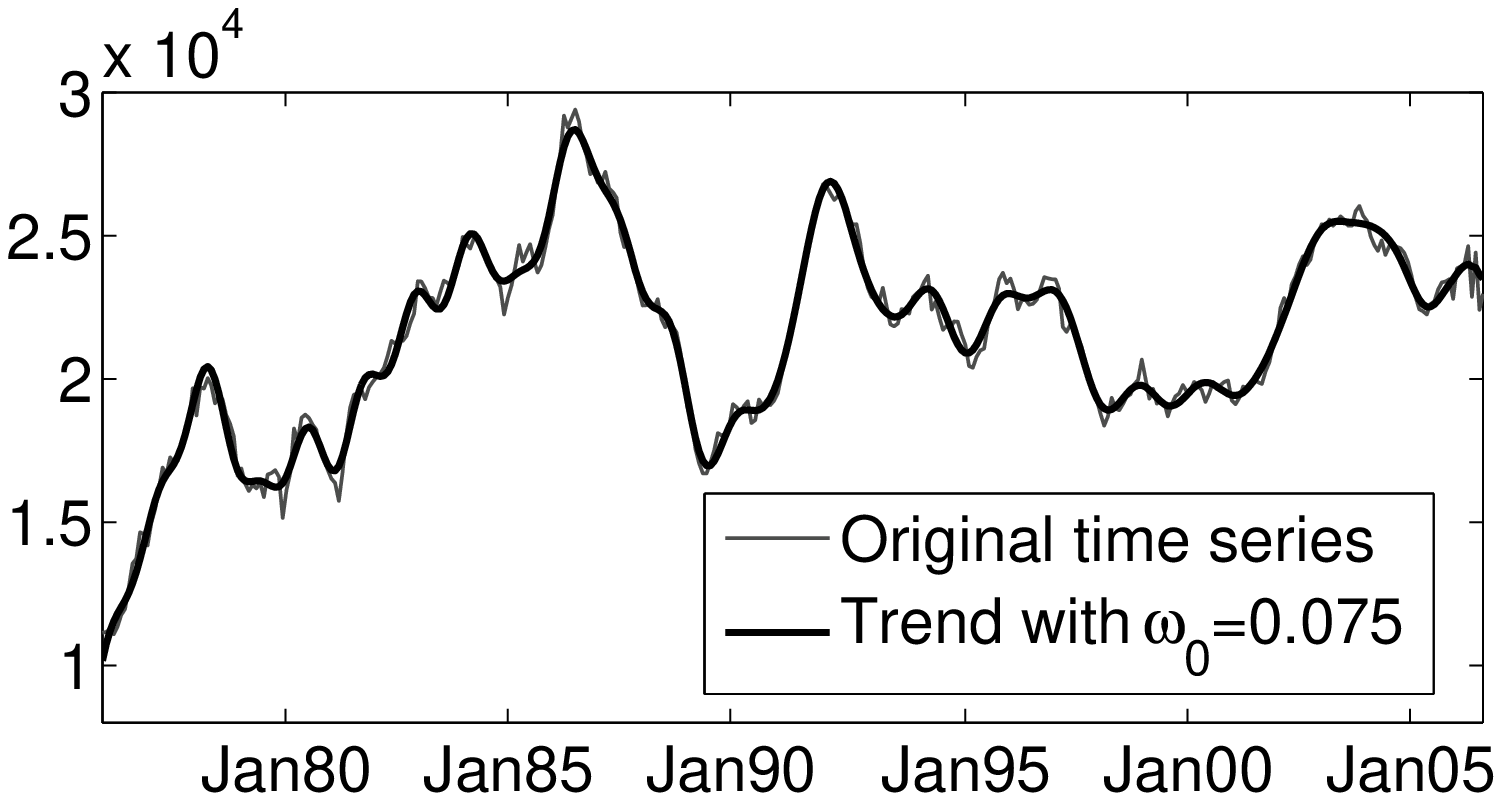}
    \includegraphics[width=.495\textwidth]{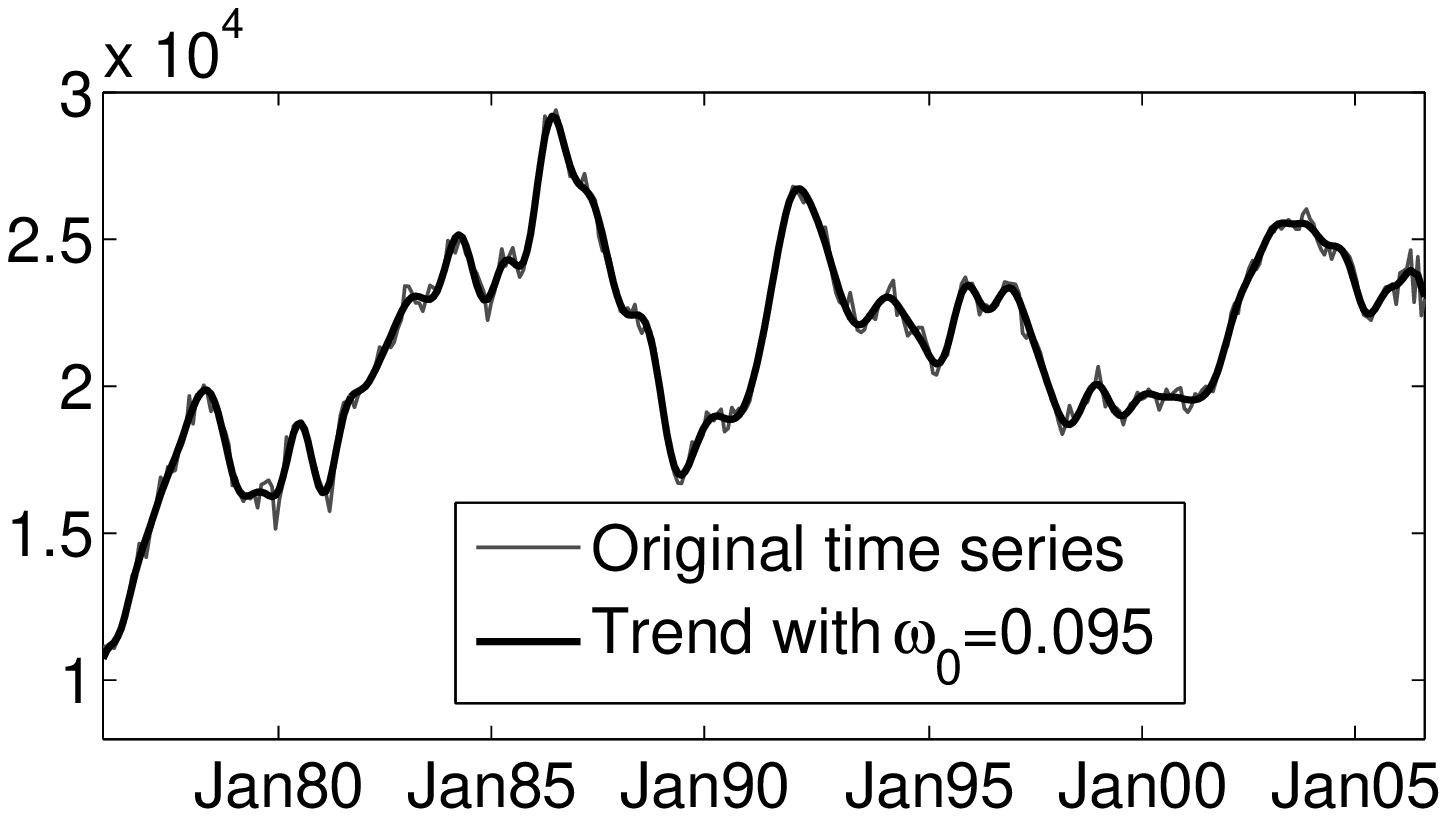}
\end{center}
  \caption{Unemployment level in Alaska: extracted trends of different scales with $\omega_0=0.01, 0.02, 0.05, 0.075$, and $0.095$
  ($L=180$, $\Delta\mathcal{C}=0.01$, and $\Delta\mathcal{R}=0.05$).}
  \label{fig:trends}
\end{figure}

We extracted trends of different scales using the following values of $\omega_0$: 
0.01, 0.02, 0.05, 0.075, and 0.095, see Figure~\ref{fig:trends} for the results.
The value $0.095 \approxeq \lceil 33/369\cdot 180 \rceil / 180$ was selected according 
to~(\ref{eq:w0_LN}), where $M_X^N \approxeq 5.19\cdot 10^5$.
The value 0.075 is the default value for monthly data (section~\ref{sec:omega0}).
Other values (0.01, 0.02 and 0.05) were considered for better illustration of how the value of $\omega_0$
influences the scale of the extracted trend. The search for $\mathcal{C}_0$ was performed
as described in section~\ref{sec:param_C0} in the interval $[0.5, 1]$ with 
the step $\Delta\mathcal{C}=0.01$ and $\Delta\mathcal{R}=0.05$.
\vspace*{5mm}
\vspace*{5mm}

\section{CONCLUSIONS}
\label{sec:conclusions}
SSA is an attractive approach to trend extraction because it: (i) requires no model
specification of time series and trend, (ii) extracts trend of noisy time series
containing oscillations of unknown period. In this paper, we presented
a method which inherits these properties and is easy to use since 
it requires selection of only two parameters.

\textit{Acknowledgments.} The author warmly thanks his Ph.D. thesis advisor Nina Golyandina for her supervision
and guidance that helped him to produce the results presented in this paper. The author
greatly appreciates an anonymous reviewer for his valuable and constructive comments.


\end{document}